\documentclass[lettersize,journal]{IEEEtran}
\usepackage{amsmath,amsfonts}
\usepackage{algorithmic}
\usepackage{algorithm}
\usepackage{array}
\usepackage[caption=false,font=normalsize,labelfont=sf,textfont=sf]{subfig}
\usepackage{textcomp}
\usepackage{stfloats}
\usepackage{url}
\usepackage{verbatim}
\usepackage{graphicx}
\usepackage{cite}
\usepackage{comment}

\usepackage{booktabs}
\newcommand{\beginsupplement}{%
        \setcounter{table}{0}
        \renewcommand{\thetable}{S\arabic{table}}%
        \setcounter{figure}{0}
        \renewcommand{\thefigure}{S\arabic{figure}}%
     }
\usepackage{adjustbox}
\usepackage{enumitem}
\usepackage{hyperref}
\begin{document}

\title{Large-scale unsupervised audio pre-training for video-to-speech synthesis}

\author{Triantafyllos Kefalas, Yannis Panagakis ,~\IEEEmembership{Member,~IEEE}, Maja Pantic ,~\IEEEmembership{Fellow,~IEEE}
        % <-this % stops a space
%\thanks{This paper was produced by the IEEE Publication Technology Group. They are in Piscataway, NJ.}% <-this % stops a space
%\thanks{Manuscript received April 19, 2021; revised August 16, 2021.}}
\thanks{This work has been submitted to the IEEE for possible publication. Copyright may be transferred without notice, after which this version may no longer be accessible. Corresponding author: Triantafyllos Kefalas (email: tk15@imperial.ac.uk)}
\thanks{Triantafyllos Kefalas and Maja Pantic are with the Department of Computing, Imperial College London, UK}
\thanks{Yannis Panagakis is with the Department of Informatics and Telecommunications, University of Athens, Greece}}

\markboth{SUBMITTED TO IEEE}%
{Shell \MakeLowercase{\textit{et al.}}: A Sample Article Using IEEEtran.cls for IEEE Journals}


\maketitle

\begin{abstract}
Video-to-speech synthesis is the task of reconstructing the speech signal from a silent video of a speaker. Most established approaches to date involve a two-step process, whereby an intermediate representation from the video, such as a spectrogram, is extracted first and then passed to a vocoder to produce the raw audio. Some recent work has focused on end-to-end synthesis, whereby the generation of raw audio and any intermediate representations is performed jointly. All such approaches involve training on data from almost exclusively audio-visual datasets, i.e. every audio sample has a corresponding video sample. This precludes the use of abundant audio-only datasets which may not have a corresponding visual modality (e.g. audiobooks, radio podcasts, speech recognition datasets etc.), as well as audio-only architectures that have been developed by the audio machine learning community over the years. In this paper we propose to train encoder-decoder models on more than 3,500 hours of audio data at 24kHz, and then use the pre-trained decoders to initialize the audio decoders for the video-to-speech synthesis task. The pre-training step uses audio samples only and does not require labels or corresponding samples from other modalities (visual, text). We demonstrate that this pre-training step improves the reconstructed speech and that it is an unexplored way to improve the quality of the generator in a cross-modal task while only requiring samples from one of the modalities. We conduct experiments using both raw audio and mel spectrograms as target outputs and benchmark our models with existing work.
\end{abstract}

\begin{IEEEkeywords}
Video-to-speech, speech synthesis, generative adversarial networks (GANs), conformer, pre-training
\end{IEEEkeywords}

\section{Introduction}
\IEEEPARstart{S}{peech} is one of the fundamental means of human communication and involves a speaker expressing an idea through language by producing a sound wave, followed by a listener perceiving and interpreting the sound wave \cite{an_overview_of_deep_learning_based_av_speech_enhancement_and_separation}. Although speech is communicated primarily through sound, humans perceive it by paying attention visual cues as well, such as facial expressions and lip movements. This natural co-ocurrence and coupling of the audio and visual speech signals has led to increasing interest by researchers in incorporating visual information in speech processing tasks. For example, the visual modality has been used to improve performance in in speech enhancement \cite{the_coversation_deep_av_speech_enhancement, towards_next_generation_lipreading_driven_hearing_aids, lip_reading_driven_deep_learning_approach_for_speech_enhancement, my_lips_are_concealed, lite_av_speech_enhancement}, speech separation \cite{multi_modal_multi_target_speech_separation, looking_to_listen_at_the_cocktail_party, visualvoice, deep_av_speech_separation_with_attention_mechanism}, and speech recognition \cite{recent_advances_in_the_automatic_recognition_of_av_speech, deep_av_speech_recognition, rnn_transducer_for_av_speech_recognition, visual_features_for_context_aware_speech_recognition, end_to_end_av_speech_recognition, av_speech_recognition_with_hybrid_ctc_attention_architecture, end_to_end_av_speech_recognition_with_conformers}. In particular, the visual modality provides complementary information when the input speech signal is noisy or corrupted, given that the acoustic noise may be independent of the visual stream, as with background noise for example.

\begin{figure}
  \includegraphics[width=\linewidth]{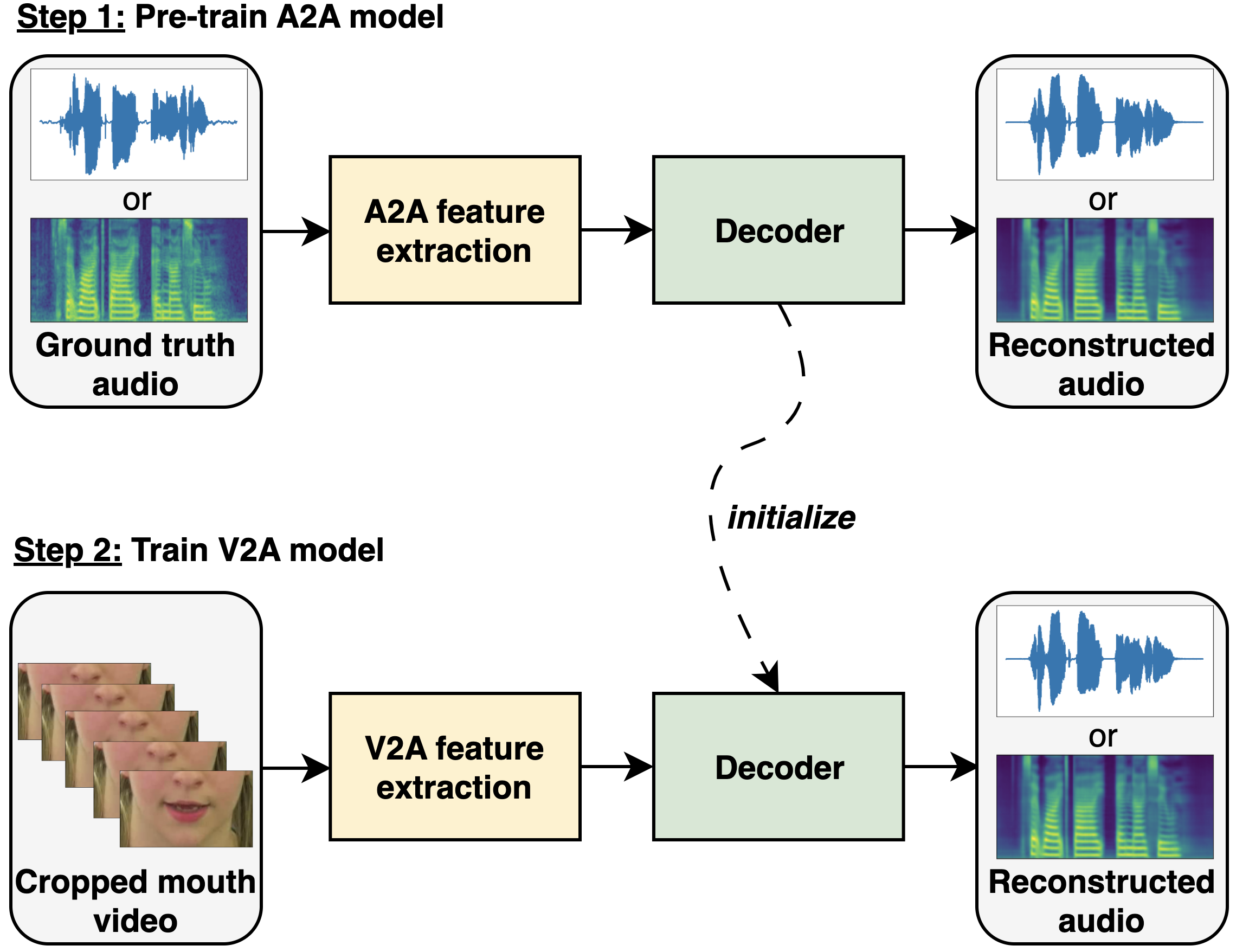}
\caption{High-level overview of the proposed method}
\label{high_level_v2a_and_a2a}
\end{figure}

However, if we do not have access to the input speech signal or if it has a very high level of noise or corruption, speech processing using the audio modality only is infeasible. This motivates the need to reconstruct the underlying speech signal using the visual modality only, leading to the task of video-to-speech synthesis. This task is useful in multiple real-world scenarios, such as speech enhancement for videoconferencing in noisy conditions \cite{ephrat2017vid2speech}, understanding surveillance silent videos\cite{ephrat2017vid2speech, av_event_recognition_in_surveillance}, generating speech for patients suffering from aphonia \cite{end_to_end_video_to_speech_synthesis_using_gans}, and making silent speech interfaces for conversational privacy and silence-required environments \cite{silent_speech_interfaces}.

A possible approach for video-to-speech synthesis is to predict the spoken words first from the video (lipreading) and then use these words to infer the missing audio (text-to-speech, TTS) \cite{end_to_end_video_to_speech_synthesis_using_gans}. Although this approach leverages the considerable recent advancements in both lipreading and TTS, it has several drawbacks as text information does not capture speaker identity characteristics (e.g. gender and age) or emotion and requires text labels during training. Furthermore, the accuracy of the lip-reading model provides an upper bound on the word accuracy of the generated sound. 

The limitations of the two-step lipreading-TTS approach have thus motivated the reconstruction of speech directly from video. Most previous works have focused on predicting intermediate audio features from video such as linear predictive coding (LPC) coefficients \cite{reconstructing_intelligible_audio_speech_from_visual_speech_features, ephrat2017vid2speech}, mel-filterbank features \cite{reconstructing_intelligible_audio_speech_from_visual_speech_features} and mel spectrograms \cite{improved_speech_reconstruction_from_silent_video, lip2audspec, lip2wav, lip_to_speech_synthesis_with_visual_context_attention_gan, speech_prediction_in_silent_videos_using_vaes, speaker_disentanglement_in_video_to-speech_conversion, svts}, followed by reconstruction of the raw waveform with a vocoder. There has also been some recent work on direct generation of the raw waveform without predicting intermediate audio features \cite{video_driven_speech_reconstruction_using_gans, end_to_end_video_to_speech_synthesis_using_gans}.

In particular some recent work \cite{lipsound2, svts} has investigated leveraging the co-occurrence of audio and video, and the abundance of unlabelled audio-visual data, to either pretrain \cite{lipsound2} video-to-speech models on large-scale audio-visual datasets and then fine-tune them on the datasets of interest, or train jointly on both the dataset of interest and an additional large dataset \cite{svts}. Although empirical results show that both pre-training and co-training improve the quality of reconstructed speech, they both require an auxiliary audio-visual dataset, i.e. that every audio sample has a corresponding video sample. This formulation prevents the use of the large number of audio-only datasets and the abundance of audio-only data which may not have a corresponding visual stream. This includes speech recognition datasets\cite{common_voice, spoken_wikipedia, voxpopuli, gigaspeech}, text-to-speech datasets \cite{libri_tts, hifi_tts, vctk, omceb, ljspeech}, audiobooks \cite{librispeech}, podcasts \cite{spotify_podcasts_dataset} and radio shows \cite{boston_university_radio_news_corpus}.

In this work we investigate pre-training the audio decoder of a video-to-speech synthesis (V2A) model on large volumes of audio-only data and then fine-tuning it on the video-to-speech task. We build upon the models of \cite{end_to_end_video_to_speech_synthesis_using_gans} and \cite{svts}, which generate raw waveform and mel spectrograms respectively, by proposing architectural changes as well as pre-training procedures.

In the raw waveform model inspired by \cite{end_to_end_video_to_speech_synthesis_using_gans} we replaced the Wasserstein GAN \cite{wasserstein_gan} with a Least Squares GAN (LS-GAN) \cite{lsgan} with a waveform discriminator for faster training. Furthermore, we added residual stacks, containing dilated convolutions, after each convolution layer of the decoder to increase its receptive field. These changes are inspired by the MelGAN architecture \cite{melgan} which has been the basis for state of the art neural vocoders \cite{melgan, multi_band_melgan, style_melgan, hifi_gan}, and by Parallel WaveGAN\cite{parallel_wavegan}. We also replaced the PASE and log-spectrogram losses \cite{pase, end_to_end_video_to_speech_synthesis_using_gans} with the multi-resolution STFT loss from \cite{parallel_wavegan, multi_band_melgan}. Finally, we include a pre-trained face (speaker) encoder to extract a face (speech) identity vector respectively as a conditioning variable, which is then combined with the visual features and fed into a bidirectional LSTM \cite{bidirectional_lstm} as input to the audio decoder.

In the mel spectrogram model inspired by \cite{svts} we insert a temporal module before the audio decoder, containing a bidirectional LSTM followed by a temporal upsampling operation \cite{autovc} and another bidirectional LSTM. The temporal upsampling operation increases the number of timesteps of an input and matches the lower sampling rate of video with the higher sampling rate of the corresponding mel spectrogram. Finally, we include face (speaker) identity vectors in the input to the temporal module as with the raw waveform model above.

Furthermore we introduce pre-training procedures with audio-only data by constructing a corresponding audio-to-audio (A2A) model for each of the two V2A models above. This involves replacing the part of the model prior to the decoder, in each case, with an audio encoder mirroring the decoder's structure. The loss functions are kept the same with the V2A models. The audio-to-audio models are then trained on a large volume of speech data, and the learned parameters of the audio decoders are then used to initialize the audio decoders of the V2A models.

Our contributions in this work are summarized as follows:
\begin{itemize}
\item We propose two new encoder-decoder models that reconstruct speech from a silent video (video-to-audio models). The first model consists of an encoder-decoder GAN that generates raw waveforms directly from video. It employs a Discriminator on the raw waveforms and is trained using the LS-GAN formulation \cite{lsgan} and two reconstruction losses. The second model is an encoder-decoder network that outputs mel spectrograms, which are then fed to a pre-trained neural vocoder, and employs one reconstruction loss.

\item For each of the two models above, we create a corresponding audio-to-audio model where we keep the corresponding decoder but replace the part prior to the decoder with an audio encoder that mirrors the decoder's structure. For the raw waveform GAN formulation, the encoder receives raw waveform as input and for the mel spectrogram formulation the input is a mel spectrogram.

\item We pre-train the audio-to-audio models on a combination of speech corpora in an unsupervised manner and use these pre-trained models to initialize and fine-tune the corresponding video-to-audio models.

\item We propose a modification to the batch normalization \cite{batch_norm} module which enables cross-modal parameter fine-tuning by keeping track of separate running statistics for the audio and video modalities. We apply this to the batch normalization layers of the decoder.

\item We conduct experiments on popular audio-visual datasets on seen (GRID \cite{grid_database}, TCD-TIMIT \cite{tcd_timit_databse}) and unseen (GRID\cite{grid_database}, LRW\cite{lip_reading_in_the_wild}) speakers. We compare our encoder-decoder models trained from scratch with those trained with fine-tuning, and thus measure the effect of unsupervised audio pre-training on speech quality and intelligibility.

\end{itemize}

\section{Related work}

\subsection{Lipreading}
Visual speech recognition, also known as lipreading, is the task of predicting the words of a speaker from silent videos. The idea of using lip movements to infer speech content, and particularly for noisy environments, was first proposed by \cite{visual_contribution_to_speech_intelligibility_in_noise} with initial computer experiments in audio-visual speech recognition and lipreading conducted by \cite{automatic_lipreading_to_enhance_speech_recognition} and \cite{continuous_optical_automatic_speech_recognition_by_lipreading} respectively.

Traditional lipreading methods employ handcrafted visual features, such as DCT features \cite{information_theoretic_feature_extraction_for_av_speech_recognition}, DWT features \cite{an_image_transform_approach_for_hmm_based_automatic_lipreading}, geometric features \cite{lip_feature_extraction_and_reduction_for_hmm, lip_feature_extraction_based_on_improved_jumping_snake_model, profile_view_lip_reading} as well as Active Appearance Models \cite{active_appearance_models, view_independent_computer_lip_reading, insights_into_machine_lip_reading}. These are then fed into Hidden Markov Models (HMMs) \cite{information_theoretic_feature_extraction_for_av_speech_recognition} or Support Vector Machines (SVMs) \cite{lipreading_with_local_spatiotemporal_descriptors} to predict text. More recently, research has shifted to end-to-end deep learning (DL) models which have outperformed traditional methods. Deep learning-based lipreading initially focused on word and phoneme-level prediction by separately extracting deep features and then training a classifier \cite{lipreading_using_convolutional_neural_network, deep_learning_of_mouth_shapes_for_sign_language, improved_speaker_independent_lip_reading, av_speech_recognition_using_bimodal_trained_bottleneck_features, deep_complementary_bottleneck_features_for_visual_speech_recognition}. The first end-to-end model at sentence level prediction was \cite{lipnet} and consisted of a spatiotemporal convolutional encoder followed by a bidirectional GRU \cite{empirical_evaluation_of_gated_rnn_on_sequence_modeling} and a final linear layer. The model was trained with a CTC loss \cite{connectionist_temporal_classification} and achieved state-of-the-art results on the GRID dataset \cite{grid_database}.

Several end-to-end lipreading works have followed at both the word and character level. Petridis et al. \cite{end_to_end_visual_speech_recognition_with_lstms, end_to_end_visual_speech_recognition_for_small_scale_datasets} presented a fully-connected model with LSTMs predicting words from frames of mouth regions and difference images. In \cite{lip_reading_in_the_wild} a CNN based model was applied to single-word silent videos recorded in the wild. A CNN+LSTM based phoneme prediction model followed by word decoding was employed in \cite{large_scale_visual_speech_recognition}, while \cite{combining_residual_networks_with_lstms_for_lipreading} presented a word prediction model with residual connections and LSTMs.

Other works incorporate the audio modality as well to perform audio-visual speech recognition. In \cite{end_to_end_av_speech_recognition} the mouth frames and raw audio are encoded by two ResNet-based encoders followed by bidirectional GRUs, whereas Afouras et al. \cite{deep_av_speech_recognition} employ a ResNet+Transformer back-end. We refer the reader to lip reading surveys \cite{deep_learning_based_automated_lipreading_a_survey, review_on_research_progress_of_machine_lipreading} for coverage of additional methods and approaches.

\subsection{Video-to-speech synthesis}
The majority of the literature on video-to-speech synthesis studies approaches that generate intermediate acoustic representations (e.g. mel spectrograms) followed by a separate waveform reconstruction module. In recent years some end-to-end methods (i.e. generating raw waveforms directly from the input visual stream) have been introduced as well.

In \cite{reconstructing_intelligible_audio_speech_from_visual_speech_features} the authors propose a model that receives handcrafted visual features (e.g. 2D-DCT) to predict spectral envelope representations (LPCs or mel-filterbank amplitudes) using Gaussian Mixture Models (GMMs) or fully-connected neural networks. The output was then fed into a STRAIGHT vocoder \cite{straight_vocoder} to synthesize the raw waveforms. This work was extended in \cite{generating_intelligible_audio_speech_from_visual_speech} which employed a classification framework involving the prediction of codebook entries corresponding to audio vectors, resulting in improved speech intelligibility. In \cite{ephrat2017vid2speech} CNNs are used to learn features from raw pixels (grayscale video frames) and output line spectral pairs (LSP) features. The LSPs, along with Gaussian noise, are then fed to a source-filter speech synthesizer to generate the raw waveforms. This was then improved on in \cite{improved_speech_reconstruction_from_silent_video} by employing two ResNet encoders, one for raw video and another for optical flow. The two sets of embeddings are then concatenated and fed into a fully-connected network to generate mel spectrograms, followed by a post-processing module to generate linear spectrograms and the Griffin-Lim algorithm (GLA) \cite{griffin_lim} to extract the final raw audio. Lip2AudSpec \cite{lip2audspec} initially trains a fully-connected auto-encoder network on spectrograms and then uses the bottleneck features as training targets for a CNN+RNN lip reading network. In \cite{vocoder_based_speech_synthesis} a multi-task model was presented that predicts the spectral envelope, aperiodic parameters and the fundamental frequency as inputs to a vocoder to synthesize the raw waveform. The model also performs lip reading jointly with a connectionist temporal classification (CTC) \cite{connectionist_temporal_classification}. Lip2Wav \cite{lip2wav} generates mel spectrograms from video frames using an encoder-decoder architecture involving a stack of 3-D convolutions followed by an attention-based decoder based on Tacotron2 \cite{tacotron2}. GLA is then used to decode the mel spectrograms to raw waveforms. VCA-GAN \cite{lip_to_speech_synthesis_with_visual_context_attention_gan} incorporates the entire video sequence as a global conditioning variable and employs a multi-scale generator with residual blocks, synthesizing mel spectrograms from coarse to fine-level, followed by a post-net similar to \cite{improved_speech_reconstruction_from_silent_video} and GLA to generate the waveforms. It is trained adversarially with a multi-scale discriminator as well as reconstruction and synchronization losses.

A VAE-based approach\cite{vae_paper}\cite{speech_prediction_in_silent_videos_using_vaes} was also proposed where the goal was to reconstruct speech from a silent video by modelling the uncertainty in speech generation. It employs an encoder-decoder generator with LSTMs in the bottleneck layers such that the uncertainty is modelled autoregressively as conditional probability distributions. The generator outputs mel spectrograms, with the raw waveform obtained using GLA.

In \cite{speech_reconstruction_visual_voice_memory} a key-value memory structure is used to map visual features (keys) to audio features (values). During training, visual and speech encoders extract features from the video frames and the ground truth mel spectrogram respectively and the decoder reconstructs the mel spectrogram from these concatenated features directly, or from the visual features and the predicted audio features via an intermediate memory module. During inference, the audio encoder is discarded and the visual features are fed to the memory module to produce imprinted audio features. These are then concatenated and fed to the decoder.

Recently, \cite{lipsound2} investigated pre-training a video-to-speech and a speech-to-text model on VoxCeleb2 \cite{voxceleb2_paper} and LibriSpeech \cite{librispeech} respectively, and then fine-tuning the video-to-speech model on a dataset of interest and the speech-to-text model on the predicted audio intended for lip reading. The video-to-speech model has an encoder-decoder structure with an auto-regressive bridge module in between. Teacher forcing is employed during training, where the ground truth mel spectrogram is fed to the bridge module (in addition to the visual features). During inference, predicted outputs from previous timesteps are used instead. This approach achieved state-of-the-art results in GRID and TCD-TIMIT.

SVTS \cite{svts} was recently introduced aiming to provide a V2A framework scalable to large datasets. It consists of a 3D Convolution+ResNet-18 encoder followed by a conformer decoder that generates mel spectrograms and a pre-trained neural vocoder. It achieved state-of-the-art results on the GRID and LRW \cite{lip_reading_in_the_wild} datasets and is the first approach that experimented on LRS3 \cite{lrs3}.

Finally, two recent works have utilized GANs to reconstruct raw waveforms from an input video directly. These are trained end-to-end, without predicting intermediate representations such as spectrograms. The first such work \cite{video_driven_speech_reconstruction_using_gans} proposed an encoder-decoder GAN, with a generator that receives a silent video and outputs a raw waveform. The generator consists of a convolutional encoder, followed by a GRU and a sequence of transposed convolutions in the decoder. The model is trained with three reconstruction losses and the Wasserstein GAN loss \cite{wasserstein_gan}, along with a convolutional waveform critic. This was followed by \cite{end_to_end_video_to_speech_synthesis_using_gans}, which proposed architectural changes to \cite{video_driven_speech_reconstruction_using_gans} as well as new reconstruction losses. In particular, it replaced the convolutional encoder with a 3D Convolution+ResNet-18 and proposed using a both a waveform critic and a critic for spectrograms.
\subsection{Large-scale pre-training with speech data}
In automatic speech recognition (ASR) it is a well-established practice to pre-train models on large speech corpora and then fine-tune them on a dataset/task of interest. Recent work includes  \cite{toward_domain_invariant_speech_recognition}, where the authors address the domain mismatch problem in ASR by pre-training on 162,000 hours of speech data from multiple domains with varying sampling rates, codecs and sources of noise. In SpeechStew \cite{speech_stew} a combination of publicly available speech recognition datasets is used to train a 100 million and a 1 billion parameter model using a conformer architecture \cite{conformer}.

Some recent research has explored large-scale pre-training for speech synthesis tasks. For example, the SUPERB-SG \cite{superb_sg} benchmark was introduced to evaluate pre-trained models on various tasks including speech enhancement and voice conversion. Prior work on pre-training generative models of speech has focused on learning representations for downstream classification tasks, rather than synthesis \cite{hubert}. A major challenge is that many publicly available datasets are noisy, making the synthesis of clean speech non-trivial.

To address the problem of noise in the training data, most approaches either: (1) exclude noisy subsets by manual inspection or by setting a required signal-to-noise ratio (SNR) as a threshold \cite{libri_tts, unsupervised_pre_training_for_data_efficient_tts}, or (2) modify the loss function with self-supervision to account for the noise \cite{training_speech_enhancement_systems_with_noisy_speech_datasets}. For example, \cite{unsupervised_pre_training_for_data_efficient_tts} investigates pre-training on the clean subset of LibriTTS \cite{libri_tts} on a pretext synthesis task, followed by fine-tuning the model on a low-resource language (Korean) for TTS. In \cite{training_speech_enhancement_systems_with_noisy_speech_datasets} a speech enhancement model is trained on the noisy CommonVoice \cite{common_voice} dataset using an unsupervised sound separation method during training. The proposed method results in higher quality generated speech compared to training using a traditional speech enhancement training scheme.

We note that the majority of large-scale pre-training for speech synthesis works involve pre-training the encoder/upstream module, in contrast to our work where we pre-train and fine-tune the decoder module.

\section{Video-to-audio models}
\begin{figure}
  \includegraphics[width=\linewidth]{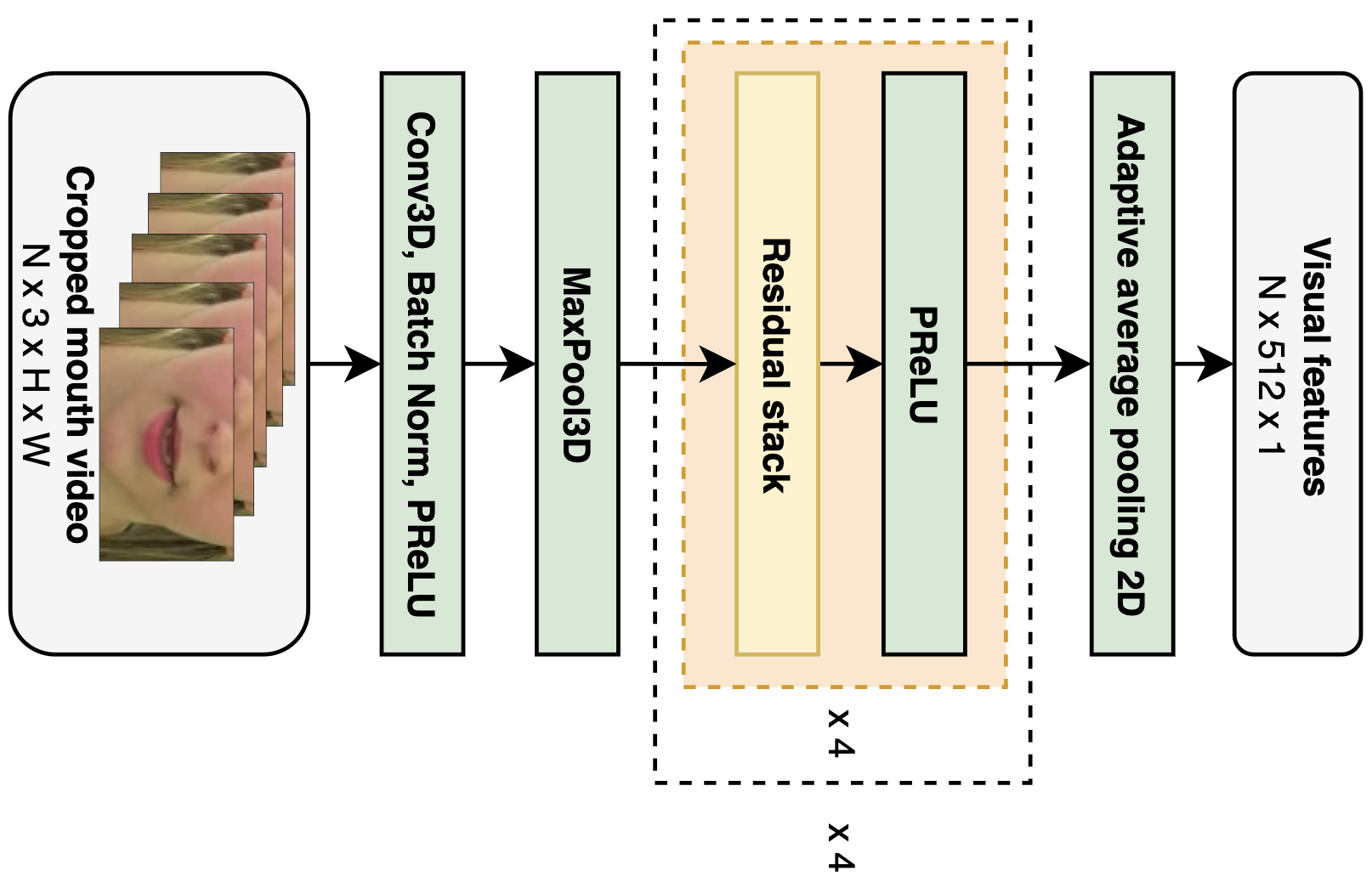} 
\caption{Video frames encoder.}
\label{video_frames_encoder_simplified}
\end{figure}

\begin{figure*}
  \includegraphics[width=\linewidth]{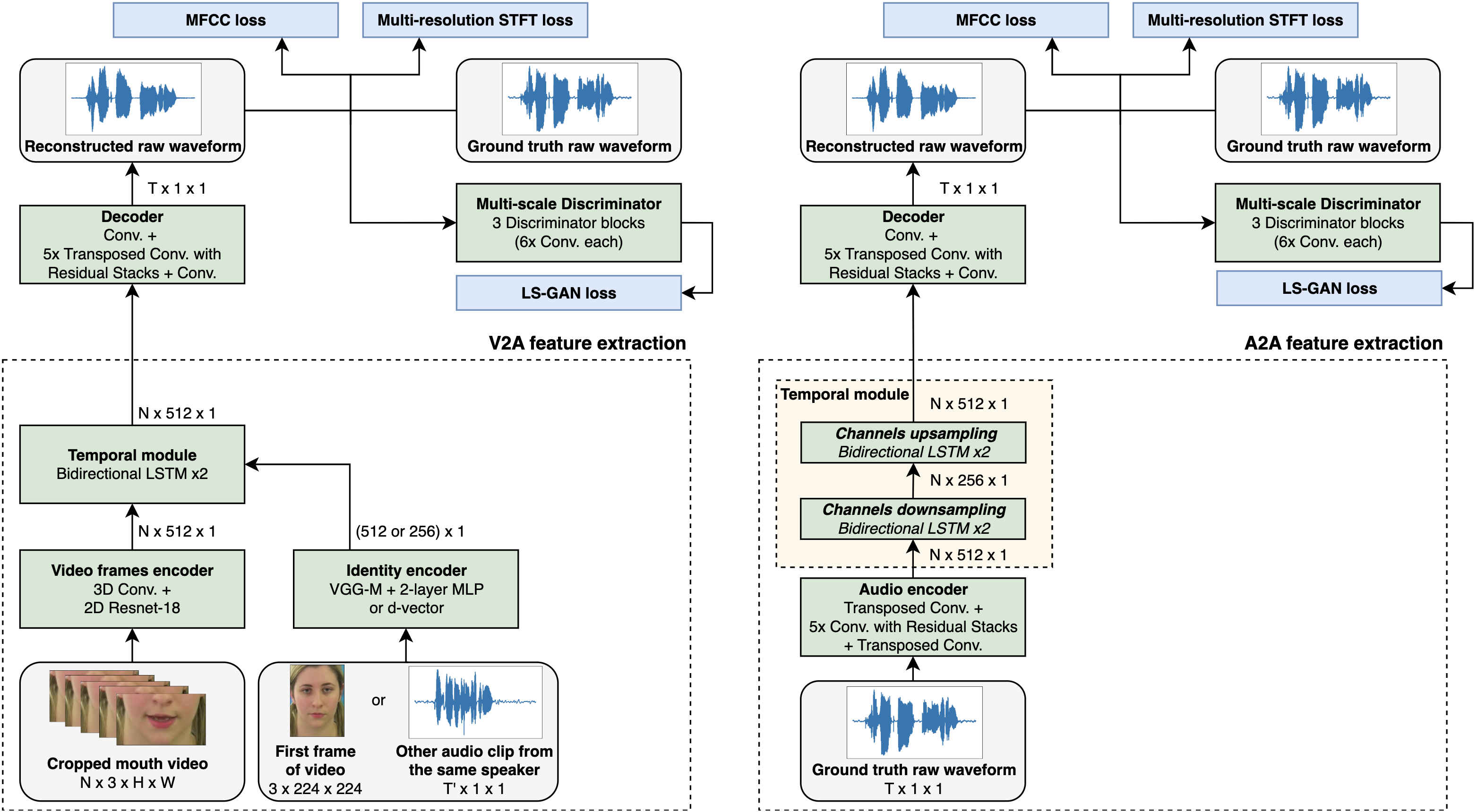}
\caption{(a) V2A-WaveGAN: Video-to-audio model with raw waveforms
\hspace{1cm} (b) A2A-WaveGAN: corresponding audio-to-audio model.}
\label{raw_waveform_high_level_v4}
\end{figure*}

\begin{figure*}
  \includegraphics[width=\linewidth]{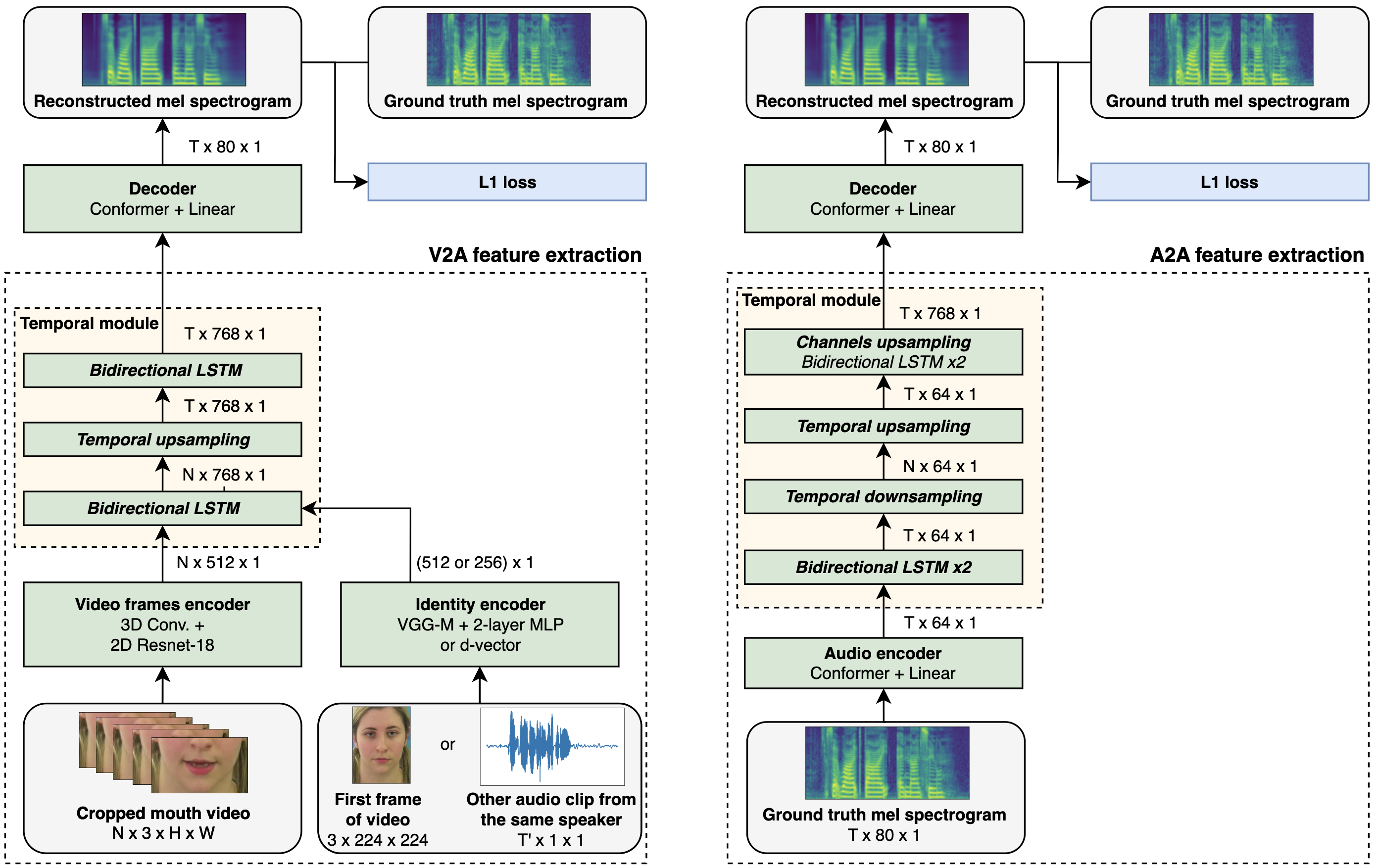}
\caption{(a) V2A-MelSpec: Video-to-audio model with mel spectrograms \hspace{1cm} (b) A2A-MelSpec: corresponding audio-to-audio model.}
\label{melspec_high_level}
\end{figure*}

Figures \ref{raw_waveform_high_level_v4}(a) and \ref{melspec_high_level}(a) show an overview of the structure of the video-to-audio models we are considering. We denote these by V2A-WaveGAN and V2A-MelSpec for the raw waveform and mel spectrogram domains respectively. In both domains, our generators consist of a video frames encoder and an identity encoder, followed by a temporal module and a decoder. The video frames encoder is based on the ResNet-18 architecture (as in \cite{deep_residual_learning_for_image_recognition, visual_speech_recognition_for_multiple_languages_in_the_wild, end_to_end_video_to_speech_synthesis_using_gans}) and the identity encoder consists of a pre-trained face (speaker) encoder trained on large-scale face (speech) data. The raw waveform generator contains the two encoders above, a bidirectional LSTM as the temporal module and a convolutional decoder. In addition, we employ a convolutional discriminator during training to improve the realism of the output waveform. The mel spectrogram generator also contains the two encoders, with an LSTM - temporal upsampling - LSTM sequence of layers as the temporal module, followed by a conformer-based decoder. We then employ a pre-trained neural vocoder\footnotemark, HiFiGAN\cite{hifi_gan}, trained on LibriTTS\cite{libri_tts}, to generate the predicted waveforms from the mel spectrograms.

\footnotetext{\texttt{\url{https://github.com/kan-bayashi/ParallelWaveGAN}}}

\subsection{Video frames encoder}

We encode the RGB video frames using a 3D spatio-temporal convolution layer, followed by batch normalization, a PReLU (Parametric Rectified Linear Unit) activation function and max pooling. This layer has a receptive field of 5 frames and is centered on the frame to be encoded, providing a temporal context of 2 future and 2 past frames. The extracted features are then passed to a 2D Resnet-18 consisting of 4 blocks with 4 convolutional residual stacks each, followed by an adaptive average pooling layer. Thus each video frame is encoded into a $512$-D representation, as shown in Fig. \ref{video_frames_encoder_simplified}.

\subsection{Identity encoder}
Inspired by multi-speaker TTS \cite{transfer_learning_from_speaker_verification}, we include an identity encoder module that encodes information about the speaker's biometric characteristics. Multi-speaker TTS, as well as some previous work on video-to-speech synthesis \cite{svts}, require audio samples to model a speaker's voice. However, we may not always have access to such audio samples, such as during inference on unseen speakers. To account for both settings of availability or no availability of audio samples, we conduct two sets of experiments with a speaker encoder and a face encoder, respectively. 

Following \cite{svts}, we encode a speaker's voice using the pre-trained speaker encoder of \cite{real_time_voice_cloning, real_time_voice_cloning_github}, trained on the task of speaker verification on VoxCeleb \cite{voxceleb1_paper}, VoxCeleb 2 \cite{voxceleb2_paper} and LibriSpeech \cite{librispeech}. Given an input video, an audio sample from the same speaker is selected at random and encoded into a $256$-D representation, known as a d-vector. The parameters of the speaker encoder are frozen during training.

In the absence of audio samples, we encode the speaker's identity by extracting an embedding from the first frame of the video. We employ a pre-trained VGG-M model \cite{return_of_devil_in_the_details} originally trained for face recognition on the VGGFace dataset \cite{deep_face_recognition, vgg_m_face_pytorch}. Given an input face image of 224x224, we extract a 4096-D embedding from the penultimate layer. This is then fed to a 2-layer feedforward network to output a 512-D final representation. The parameters of the VGG-M model are frozen during training.

Finally, given a sequence of $N$ $512$-D visual features (corresponding to $N$ frames of one video), the voice (face) embedding is concatenated at each timestep, resulting in a sequence of $768$-D features when using a speaker embedding and $1024$-D features when using a face embedding.

\subsection{Raw waveform generation}
\label{subsection:raw_waveform_generation}

\begin{figure}
  \includegraphics[width=\linewidth]{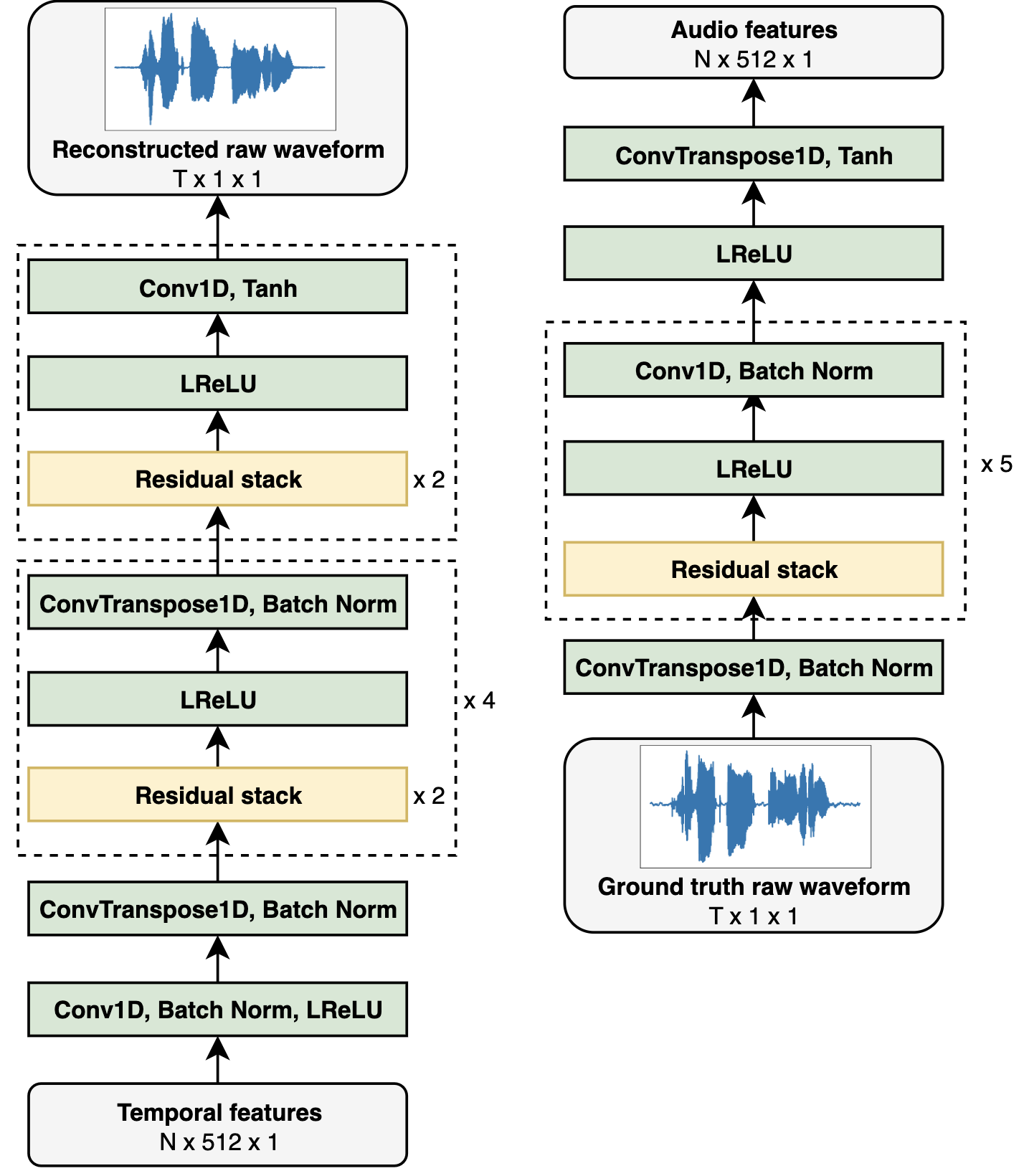}
\caption{Raw waveform decoder (left) and encoder (right).}
\label{raw_waveform_encoder_and_decoder_simplified}
\end{figure}

\begin{figure}
  \includegraphics[width=\columnwidth]{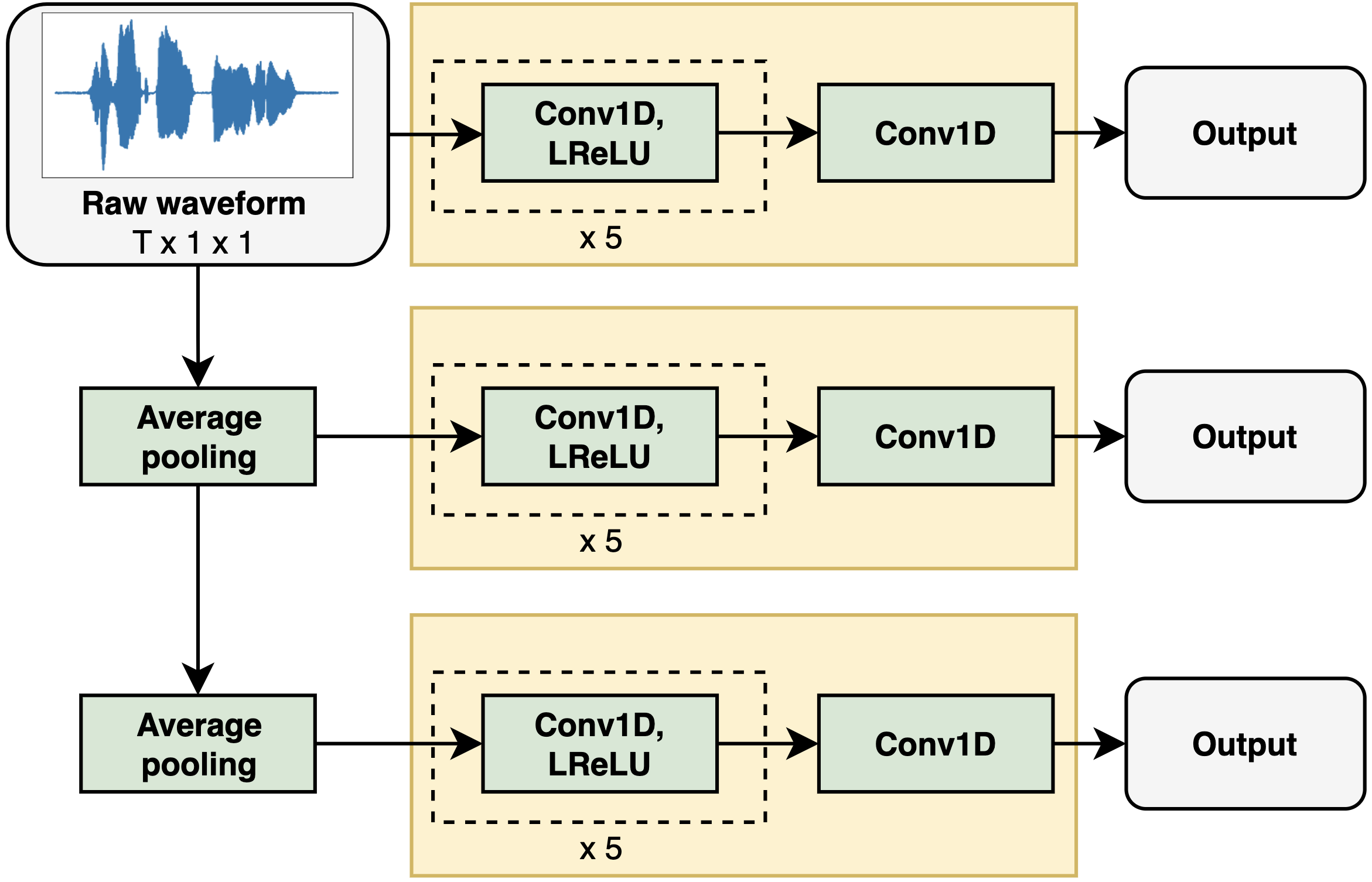}
\caption{Multi-scale discriminator for the raw waveform \cite{melgan}. Each discriminator block is shaded in yellow.}
\label{raw_waveform_discriminator}
\end{figure}

\subsubsection{Generator}
\label{subsubsection:raw_waveform_generation_generator}
Following the video and identity encoders, the resulting sequence of features is passed through the temporal module, consisting of a 2-layer bidirectional LSTM, to output a sequence of $512$-D temporal features. These are then fed as input to the decoder, which consists of a convolution and a transposed convolution block followed by 5 residual blocks and a final tanh activation function. Each residual block consists of two residual stacks followed by Leaky ReLU and a convolution/transposed convolution block (Fig. \ref{raw_waveform_encoder_and_decoder_simplified}).
The convolution blocks serve as upsampling layers, whereas the residual stacks contain dilated convolutions, with the dilation factor increasing with increasing number of stacks, to increase the induced receptive field for each output timestep. This helps place an inductive bias that the audio time-steps exhibit temporal correlation \cite{melgan}.

Given a sequence of $N$ temporal features the decoder upsamples them to a raw waveform of $T$ timesteps. Using a video frame rate of 25 frames per second and an audio sampling rate of 24 kHz the decoder thus outputs an audio segment of 960 timesteps per temporal feature.

\subsubsection{Discriminator}
\label{subsubsection:raw_waveform_generation_discriminator}

We employ the multi-scale discriminator architecture of MelGAN \cite{melgan}, shown in Fig. \ref{raw_waveform_discriminator}. This consists of 3 discriminator networks with identical architecture, each computing a low-dimensional representation of an input waveform at a given scale. The first discriminator operates at the original scale of the waveform, while we downsample by 2x for each subsequent discriminator. The rationale for using discriminators at multiple scales is that raw audio contains structures at different frequencies \cite{melgan}. Furthermore, it was found that employing a single discriminator on the raw waveform produces metallic audio \cite{melgan}. We employ weight normalization \cite{weight_normalization} for all layers of the discriminator.

\subsubsection{Loss function}
\label{subsubsection:raw_waveform_generation_loss_function}
We train the network using the LS-GAN loss \cite{lsgan}, defined as follows for the generator and the multi-scale discriminator as follows:

\begin{equation}
\label{eq:gen_adv_loss}
L_G = \mathbb{E}_{\mathbf{\Tilde{x}} \sim \mathbb{P}_G}\bigg{[}\sum_{k = 1}^{K}(D_k(\mathbf{\Tilde{x}})-1)^2\bigg{]}
\end{equation}

\begin{equation}
\label{eq:disc_adv_loss}
L_D = \mathbb{E}_{\mathbf{x} \sim \mathbb{P}_X}\bigg{[}\sum_{k = 1}^{K} (D_k(\mathbf{x})-1)^2\bigg{]} + \mathbb{E}_{\mathbf{\Tilde{x}} \sim \mathbb{P}_G}\bigg{[}\sum_{k = 1}^{K}D_k(\mathbf{\Tilde{x}})^2\bigg{]}
\end{equation}

\noindent
where $G$ is the generator, $D$ is the multi-scale discriminator, $D_k$ is the $k$th discriminator for $k=1, 2, ... K$ scales, $\mathbf{x} \sim \mathbb{P}_X$ are samples from the data distribution and $\mathbf{\Tilde{x}} \sim \mathbb{P}_G$ are samples from the generator's distribution.

We also employ two reconstruction losses to train the generator. The first is the multi-resolution STFT loss \cite{parallel_wavegan}. A single-resolution STFT loss $L_S$ is defined as:

\begin{gather}
\label{eq:single_stft_loss}
L_S(\mathbf{x}, \mathbf{\Tilde{x}}) = L_{SC}(\mathbf{x}, \mathbf{\Tilde{x}}) + L_{MAG}(\mathbf{x}, \mathbf{\Tilde{x}}) \\
L_{SC}(\mathbf{x}, \mathbf{\Tilde{x}}) = \frac{|| \ |STFT(\mathbf{x})| \ ||_F - || \ |STFT(\mathbf{\Tilde{x}})| \ ||_F}{|| \ |STFT(\mathbf{x})| \ ||_F} \\
L_{MAG}(\mathbf{x}, \mathbf{\Tilde{x}}) = \frac{1}{n}|| \ log|STFT(\mathbf{x})| - log|STFT(\Tilde{\mathbf{x}})| \ ||_1
\end{gather}

\noindent
and consists of the spectral convergence loss $L_{SC}$ and the log-STFT magnitude loss $L_{MAG}$, where $||\cdot||_F$ $||\cdot||_1$ are the Frobenius and L1 norms respectively and $n$ is the number of elements in the spectrogram.

By combining $M$ STFT losses with different analysis parameters (e.g. FFT size, window size, hop size) we obtain the multi-resolution STFT loss:

\begin{equation}
\label{eq:multi_resolution_stft_loss}
L_{MR\_ STFT}(\mathbf{x}, \mathbf{\Tilde{x}}) = \frac{1}{M}\sum_{m=1}^{M}L_S^{(m)}(\mathbf{x}, \mathbf{\Tilde{x}})
\end{equation}

\noindent
where the $m = 1, 2, ..., M$ denotes the $m$th set of STFT analysis parameters.

The second reconstruction loss is the MFCC loss, introduced in \cite{end_to_end_video_to_speech_synthesis_using_gans}, which aims to increase accuracy and intelligibility of the generated speech given that MFCCs (mel-frequency cepstral coefficients \cite{mfcc_paper}) are often used in speech and emotion recognition \cite{end_to_end_video_to_speech_synthesis_using_gans}. It is defined as:

\begin{equation}
\label{eq:mfcc_loss}
L_{MFCC}(\mathbf{x}, \mathbf{\Tilde{x}}) = ||MFCC(\mathbf{x}) - MFCC(\mathbf{\Tilde{x}})||_1
\end{equation}

\noindent
where the $MFCC$ function extracts 25 MFCCs from the raw waveform.

The final loss function of the generator combines the generator's adversarial loss with the aforementioned two reconstruction losses:

\begin{equation}
\label{eq:v2a_wavegan_generator_loss}
L_{GEN} = \lambda_1 L_G + \lambda_2 L_{MR\_ STFT} + \lambda_3 L_{MFCC}
\end{equation}

\noindent
where $\lambda_1, \lambda_2, \lambda_3 >0$ are hyperparameters.

Following \cite{end_to_end_video_to_speech_synthesis_using_gans} we tune these coefficients sequentially by incrementally finding the values that yield the lowest word error rate on the validation set of GRID (4 speakers, seen). This yields $\lambda_1 = 1.0$, $\lambda_2 = 80.0$, $\lambda_3 = 15.0$.

\subsection{Mel spectrogram generation}
\label{subsection:melspec_generation}

\begin{figure}
  \includegraphics[width=\linewidth]{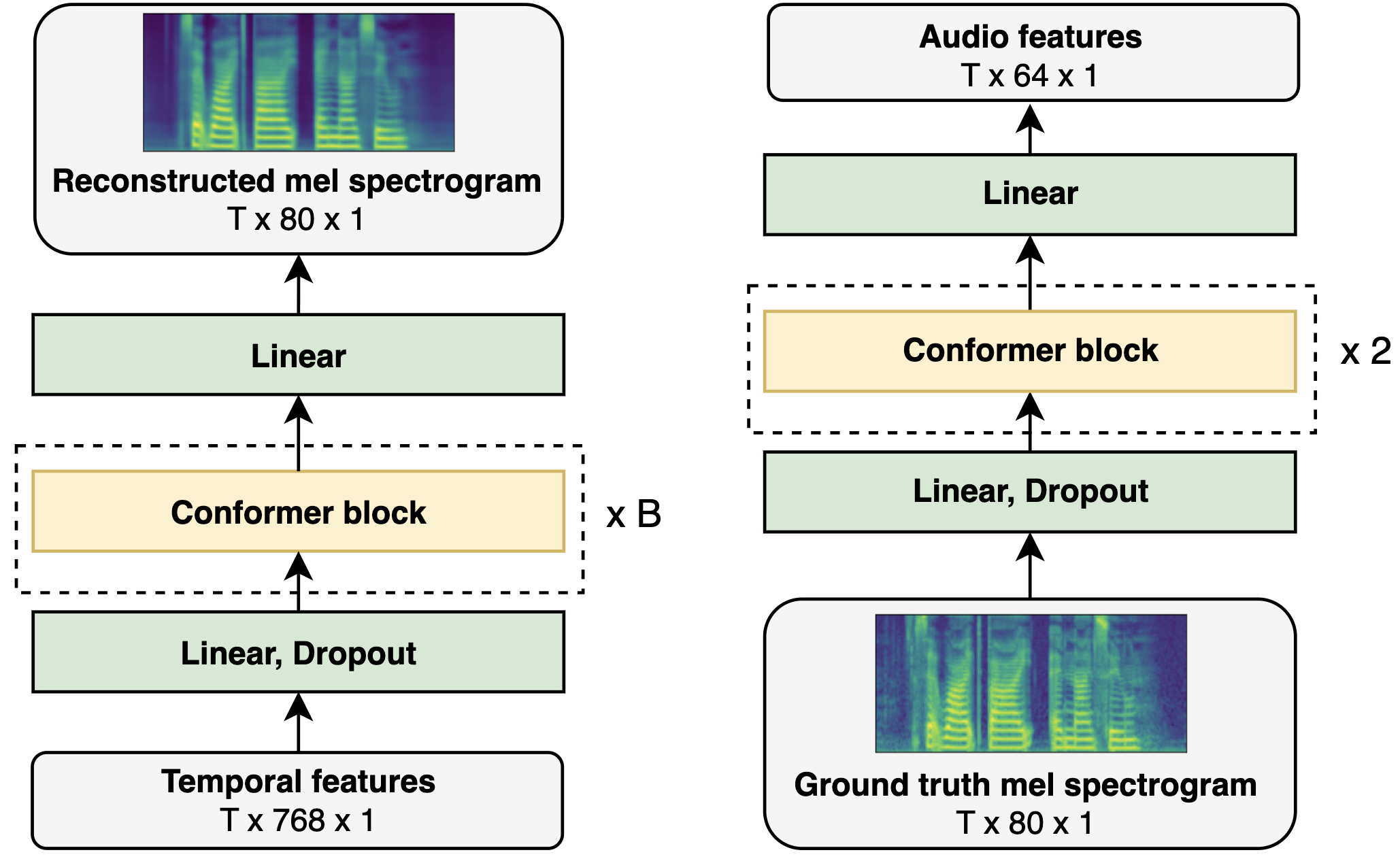}
\caption{Mel spectrogram decoder (left) and encoder (right).}
\label{melspec_encoder_and_decoder_simplified}
\end{figure}

\subsubsection{Generator}
\label{subsubsection:melspec_generation_generator}
As with raw waveform generation, we begin by feeding the sequence of features from the video and identity encoders to a temporal module. This consists of a 1-layer bidirectional LSTM followed by nearest-neighbor upsampling along the time axis and another 1-layer bidirectional LSTM, to output a sequence of $768$-D temporal features. These are then passed through the decoder which consists of a linear layer with dropout followed by $B$ conformer blocks and a final linear layer to output a mel spectrogram, as shown in Fig. \ref{melspec_encoder_and_decoder_simplified} and Table \ref{table:v2a_melspec_decoder_architectures}. The number of conformer blocks $B$ is a hyperparameter proportional to the size of the training dataset, in line with previous work \cite{svts}, resulting in three versions of the generator.

Using a video frame rate of 25 frames per second and a mel spectrogram sampling rate of 80 frames per second the decoder outputs a mel spectrogram of 3.2 timesteps per temporal feature (i.e., per video frame).

\subsubsection{Loss function}
\label{subsubsection:melspec_generation_loss_function}

We employ an L1 loss on the mel spectrograms:

\begin{equation}
\label{eq:v2a_melspec_generator_loss}
L_{GEN}(\mathbf{X}, \mathbf{\Tilde{X}}) = || \mathbf{X} - \mathbf{\Tilde{X}} ||_1
\end{equation}

\noindent
where $\mathbf{X}$ and $\mathbf{\Tilde{X}}$ are the ground truth and synthesized mel spectrograms, respectively.

\begin{table}[h]
\captionsetup{justification=centering}
\caption{Summary of V2A-MelSpec decoder architectures}
\begin{adjustbox}{width=\columnwidth}
\begin{tabular}{@{}lccc@{}}
\toprule
\multicolumn{1}{c}{Model} & \begin{tabular}[c]{@{}c@{}}V2A-MelSpec\\ (VS)\end{tabular} & \begin{tabular}[c]{@{}c@{}}V2A-MelSpec\\ (S)\end{tabular} & \begin{tabular}[c]{@{}c@{}}V2A-MelSpec\\ (M)\end{tabular} \\ \midrule
Conformer blocks          & 2                                                          & 6                                                         & 12                                                        \\
Attention dim.            & 256                                                        & 256                                                       & 256                                                       \\
Attention heads           & 4                                                          & 4                                                         & 4                                                         \\
Conv. kernel size         & 31                                                         & 31                                                        & 31                                                        \\
Feedforward dim.          & 2048                                                       & 2048                                                      & 2048                                                      \\ \bottomrule
\end{tabular}
\end{adjustbox}
\label{table:v2a_melspec_decoder_architectures}
\end{table}

\section{Audio-to-audio models}
Figures \ref{raw_waveform_high_level_v4}(b) and \ref{melspec_high_level}(b) show an overview of the structure of the proposed audio-to-audio models, denoted by by A2A-WaveGAN and A2A-MelSpec for the raw waveform and mel spectrogram domains respectively. These receive an audio input, extract low-dimensional features, and aim to reconstruct it. Our raw waveform and mel spectrogram generators both consist of an audio encoder followed by a temporal module and a decoder. The architecture of the decoder is the same as of the corresponding video-to-audio model, while the audio encoder mirrors the decoder's structure.

\subsection{Rationale for temporal module}
The temporal module acts as a bridge between the A2A and the V2A models and satisfies two constraints:

Firstly, given that we aim to use the learned parameters of the A2A decoder to initialize the decoder in the V2A task, it is necessary that the decoder receives input of the same dimensionality in both the A2A and V2A tasks. Secondly, we want the A2A models to learn useful latent representations of the data, which we accomplish by ensuring these representations have lower dimensionality than the input (as is common practice with autoencoders \cite{Goodfellow-et-al-2016}). However, the typical video has much higher resolution per unit of time than its corresponding audio stream. Hence a suitable latent representation for the A2A task may have a dimensionality that is too low for the V2A task, which would discard useful information from the input video.

To address these two constraints, we construct a temporal module for the A2A task which contains downsampling-upsampling steps across the channels and/or time axes. The downsampling steps produce compressed intermediate features of the input audio, while the upsampling steps receive these as input to produce features of higher dimensionality. Thus, the input to the decoder is of dimensionality suitable for V2A but it has been constructed from features of lower dimensionality.

\subsection{Raw waveform generation}
\subsubsection{Generator}
The audio encoder (Fig. \ref{raw_waveform_encoder_and_decoder_simplified}) consists of a transposed convolution block followed by 5 residual blocks, a Leaky ReLU activation, and a final convolution layer with a tanh activation function. The residual blocks contain one residual stack followed by Leaky ReLU and a convolution block. Note that in mirroring the structure of the decoder, we have replaced convolutions by transposed convolutions in the corresponding layers. Given a raw waveform sampled at 24 kHz the audio encoder outputs a sequence of $512$-D features at a rate of 25 features per second, to match the audio and video sampling rates used in our experiments.

The temporal module consists of two 2-layer bidirectional LSTMs which, in addition to modelling temporal interactions, downsample and upsample the input channels to $256$ and $512$ respectively.

The decoder has the same architecture as in the V2A task (\ref{subsubsection:raw_waveform_generation_generator}).

\subsubsection{Discriminator} We employ the same discriminator architecture as in the V2A task (\ref{subsubsection:raw_waveform_generation_discriminator}).

\subsubsection{Loss function} We employ the loss function defined in the V2A task (\ref{subsubsection:raw_waveform_generation_loss_function}) using the same loss coefficients.

\subsection{Mel spectrogram generation}

\subsubsection{Generator}
The audio encoder (Fig. \ref{melspec_encoder_and_decoder_simplified}) consists of a linear layer with dropout, followed by 2 conformer blocks and a final linear layer. Given an input mel spectrogram of $T$ timesteps, the audio encoder produces a $64$-D sequence of $T$ features.

The temporal module contains a 2-layer bidirectional LSTM, followed by a downsampling-upsampling operation along the time axis and a final 2-layer bidirectional LSTM that upsamples across the channel axes. We employ the downsampling-upsampling procedure of AutoVC \cite{autovc}. An input sequence of 80 features per second (corresponding to a mel spectrogram of 80 timesteps) is downsampled to a sequence of 25 features per second to match the mel spectrogram and video sampling rates used in our experiments. These are then upsampled across time to the original 80 features per second. Finally, a 2-layer bidirectional LSTM upsamples these features across channels to produce a sequence of $768$-D features.

The decoder has the same architecture as in the V2A task (Table \ref{table:v2a_melspec_decoder_architectures}) resulting in three corresponding models: A2A-MelSpec (VS), A2A-MelSpec (S) and A2A-MelSpec (M). Simiarly we use a pre-trained HiFiGAN \cite{hifi_gan} to convert the reconstructed mel spectrograms to raw waveforms.

\subsubsection{Loss function} We employ the loss function defined in the V2A task (\ref{subsubsection:melspec_generation_loss_function}).

\section{Experimental methodology and setup}
\subsection{Datasets}
\label{subsection:datasets}
\subsubsection{Audio-visual}
We train the video-to-audio models on three audio-visual datasets, widely used in the video-to-speech literature: GRID \cite{grid_database}, TCD-TIMIT \cite{tcd_timit_databse} and LRW\cite{lip_reading_in_the_wild}, summarized in Table \ref{table:av_datasets}.

GRID \cite{grid_database} consists of 33 speakers uttering structured sentences recorded in laboratory conditions, each containing 6 words chosen at random from a fixed vocabulary of 51 words. We experiment with three versions of this dataset: (1) a seen speaker setting with 4 speakers, employed in many previous works \cite{end_to_end_video_to_speech_synthesis_using_gans, video_driven_speech_reconstruction_using_gans, lip2wav, lip_to_speech_synthesis_with_visual_context_attention_gan}; (2) a seen speaker setting with 33 speakers, originally proposed in \cite{end_to_end_video_to_speech_synthesis_using_gans} and (3) an unseen speaker setting with 33 speakers used in \cite{end_to_end_video_to_speech_synthesis_using_gans, video_driven_speech_reconstruction_using_gans, lip_to_speech_synthesis_with_visual_context_attention_gan}.

TCD-TIMIT \cite{tcd_timit_databse} is composed of 62 speakers (3 lipspeakers, 59 volunteers) uttering sentences in laboratory conditions. In line with previous works \cite{lip2wav, end_to_end_video_to_speech_synthesis_using_gans, lip_to_speech_synthesis_with_visual_context_attention_gan} we use the audio and video clips of the 3 lipspeakers only, amounting to 377 clips per lipspeaker. We employ a seen speaker split proposed in \cite{end_to_end_video_to_speech_synthesis_using_gans}.

LRW \cite{lip_reading_in_the_wild} contains audio-visual clips of one word utterances from hundreds of speakers, with a vocabulary of 500 words, recorded 'in-the-wild'. These clips were sourced from television shows and feature variations in background noise, head pose and lighting conditions. Thus, this dataset is more challenging for speech reconstruction tasks than GRID and TCD-TIMIT.

\begin{table}[h]
\captionsetup{justification=centering}
\caption{Summary of audio-visual datasets used for V2A}
\begin{adjustbox}{width=\columnwidth}
\begin{tabular}{@{}lccc@{}}
\toprule
\textbf{Dataset}                & \textbf{\begin{tabular}[c]{@{}c@{}}Training set\\ (samples/hours)\end{tabular}} & \textbf{\begin{tabular}[c]{@{}c@{}}Validation set\\ (samples/hours)\end{tabular}} & \textbf{\begin{tabular}[c]{@{}c@{}}Test set\\ (samples/hours)\end{tabular}} \\ \midrule
GRID (4 speakers, seen)         & 3543 / 2.95                                                                              & 209 / 0.17                                                                                 & 209 / 0.17                                                                           \\
GRID (33 speakers, seen)        & 29333 / 24.44                                                                              & 1627 / 1.36                                                                                 & 1630 / 1.36                                                                          \\
GRID (33 speakers, unseen)      & 15648 / 13.04                                                                              & 6996 / 5.83                                                                                 & 9946 / 8.29                                                                          \\
TCD-TIMIT (3 lipspeakers, seen) & 1014 / 1.64                                                                              & 57 / 0.09                                                                                 & 60 / 0.10                                                                           \\
LRW (unseen)                    & 480456 / 154.81                                                                              & 24728 / 7.97                                                                                & 24663 / 7.95                                                                          \\ \bottomrule
\end{tabular}
\end{adjustbox}
\label{table:av_datasets}
\end{table}

\subsubsection{Audio only}

We train the audio-to-audio models on a combination of publicly available audio datasets, originally created to train speech recognition and TTS models. The datasets are in English and we use samples with a minimum sampling rate of 24kHz. We set aside the LibriTTS \texttt{dev} and \texttt{test} subsets, containing approximately 30 hours of data, as the validation set and use all remaining samples for training.

CommonVoice \cite{common_voice} is a crowd-sourced speech dataset containing over 9,000 hours of speech clips from more than 60 languages \cite{training_speech_enhancement_systems_with_noisy_speech_datasets}. It is an ongoing project with more speech clips, languages and speakers added over time. The samples feature a variety of speaker accents and recording environments, including lab conditions as well as environments with varying sources of noise. We use version 9.0 of the English subset.

LibriTTS \cite{libri_tts} and HiFiTTS \cite{hifi_tts} contain more than 800 hours of samples from more than 2,000 speakers reading audiobooks. They are both subsets of the LibriSpeech \cite{librispeech} dataset, selected for their high signal-to-noise ratios.

Spoken Wikipedia \cite{spoken_wikipedia} contains approximately 1,000 hours of speech data in English, German and Dutch, consisting of volunteers reading Wikipedia articles. We use version 2.0 of the English subset.

VCTK \cite{vctk} contains 44 hours of clean speech from 109 speakers, where each speaker reads out sentences selected from newspapers.

Finally, OMCEB \cite{omceb} contains 31 hours of clean speech from volunteers with different accents across the British Isles. The speakers read sentences from various texts in the public domain, including Wikipedia, as well as manually created sentences to highlight certain words. 

\begin{table*}[]
\centering
\caption{Speech datasets used for pre-training the A2A models}
\begin{tabular}{@{}lllll@{}}
\toprule
Dataset                  & Domain                                                                             & \begin{tabular}[c]{@{}l@{}}Sampling rate (kHz)\end{tabular} & Hours          & Number of speakers   \\ \midrule
CommonVoice Corpus 9.0 English \cite{common_voice}                & Reading sentences                                                                         & 24 - 48                                                            & 2,224            & 81,085                \\
LibriTTS \cite{libri_tts}                & Audiobooks                                                                         & 24                                                            & 586            & 2,456                \\
Spoken Wikipedia 2.0 English \cite{spoken_wikipedia} & Reading                                                                            & 8 - 96                                                              & 395            & 395                  \\
HiFiTTS \cite{hifi_tts}                 & Audiobooks                                                                         & 44.1                                                          & 292            & 10                   \\
VCTK \cite{vctk}                     & \begin{tabular}[c]{@{}l@{}}Reading sentences \end{tabular} & 48                                                            & 44             & 109                  \\
OMCEB \cite{omceb}                   & Reading sentences                                                                           & 48                                                            & 31             & 120                  \\ \midrule
\textbf{Total}           & \textbf{}                                                                          & \textbf{}                                                     & \textbf{3,572} & \textbf{$>$ 81,000} \\ \bottomrule
\end{tabular}
\label{table:audio_datasets}
\end{table*}

\subsection{Data pre-processing}

We crop the mouth from every frame of the video and feed the resulting sequence to the video frames encoder. To do this we first perform face detection using the S$^3$FD \cite{s3fd}, followed by 68-point landmark localization using a pre-trained 2D-FAN \cite{fan} and alignment of each face to a reference mean face. Finally, we extract $128\times74$ regions for GRID and $96\times96$ regions for TCD-TIMIT and LRW and normalize the resulting images. As data augmentation during training, we also apply horizontal flipping to each input frame with a 50\% probability.

All audio is sampled at 24 kHz. We extract log-mel spectrograms using 80 mel bands, an FFT size of 2048, a hop size of 12.5 ms, a window length of 50 ms, and a Hann window. In addition we clip values outside the $[-6, 6]$ range and rescale the resulting spectrogram to $[-1, 1]$.

We do not perform any pre-processing on the raw waveforms.

\subsection{Raw waveform model training and pre-training}
\label{subsection:raw_waveform_model_training_and_pretraining}

In both A2A and V2A experiments with raw waveforms we train the generator and the discriminator using a batch size of 4 and with the Adam \cite{adam} optimizer using a learning rate of $1 \times 10^{-4}$, $\beta_1 = 0.5$ and $\beta_2 = 0.99$. In line with \cite{end_to_end_video_to_speech_synthesis_using_gans} we feed a 1 second audio clip to the discriminator sampled at random from the ground truth and reconstructed audio. The generator receives input samples of duration up to 3 seconds.

\subsubsection{A2A-WaveGAN pre-training}

In our preliminary experiments, we noticed that including the CommonVoice dataset in the training data of an A2A model led to higher validation loss than without it, in line with similar observations in the TTS literature \cite{can_we_use_common_voice_to_train}. Thus we trained the A2A model in two steps: firstly, we train it on all the audio datasets in Table \ref{table:audio_datasets} (excluding CommonVoice) until the validation loss is minimized; then we fine-tune the A2A model with the lowest validation loss on all the speech datasets (including CommonVoice) with the same learning rate. We select the model with the lowest validation loss as the final A2A model.

\begin{algorithm}
\caption{V2A-WaveGAN training with alternating fine-tuning}
\begin{algorithmic}[1]
\STATE \textbf{Input:} $E_A, T_A, F_A$, \texttt{num\_epochs}, \texttt{train\_data}
\STATE \textbackslash\textbackslash \: Initialize V2A modules
\STATE $F_V \gets F_A$ 
\STATE $E_V, E_I, T_V, D_V \gets$ \texttt{random\_init}
\STATE \texttt{for epoch in num\_epochs:}
\STATE \quad \texttt{for batch in train\_data:}
\STATE \qquad \textbackslash\textbackslash \:Video frames, audio, input to identity encoder
\STATE \qquad $\mathbf{x_v}, \mathbf{x}, \mathbf{x_I} \gets $ \texttt{batch}
\STATE
\STATE \qquad \textbackslash\textbackslash \:Reconstruct speech from V2A feature extraction
\STATE \qquad $\mathbf{z_v}, \mathbf{z_I} = E_V(\mathbf{x_v}), E_I(\mathbf{x_I})$
\STATE \qquad $\mathbf{z} = T_V(\mathbf{z_v}, \mathbf{z_I})$
\STATE \qquad $\mathbf{\Tilde{x}} = F_V(\mathbf{z})$
\STATE \qquad \texttt{compute\_loss} $(\mathbf{x}, \mathbf{\Tilde{x}})$
\STATE \qquad \texttt{backpropagation} ($E_V, T_V, F_V, D_V$)
\STATE
\STATE \qquad \textbackslash\textbackslash \:Reconstruct speech from A2A feature extraction
\STATE \qquad $\mathbf{z} = T_A(E_A(\mathbf{x}))$
\STATE \qquad $\mathbf{\Tilde{x}} = F_V(\mathbf{z})$
\STATE \qquad \texttt{compute\_loss} $(\mathbf{x}, \mathbf{\Tilde{x}})$
\STATE \qquad \texttt{backpropagation} ($E_A, T_A, F_V, D_V$)
\STATE \textbf{Output: $E_A, T_A, E_V, E_I, T_V, F_V, D_V$}
\end{algorithmic}
\label{algorithm:alternating_fine_tuning}
\end{algorithm}

\subsubsection{V2A-WaveGAN training and finetuning} We conduct three sets of experiments with V2A-WaveGAN:
\label{subsubsection:v2a_wavegan_training_and_finetuning}

\begin{itemize}
\item \textbf{From scratch:} Random initialization of all parameters
\item \textbf{Basic fine-tuning:} Initializing the decoder and discriminator with the pre-trained parameters of A2A-WaveGAN. Optionally, initializing the respective parameters of the Adam optimizer with their pre-trained values.
\item \textbf{Alternating fine-tuning:} Initializing the decoder with the pre-trained parameters of A2A-WaveGAN. During training, we alternate between reconstructing speech from video frames (and the identity input) vs. reconstructing speech from the ground truth waveform. This is shown in Algorithm \ref{algorithm:alternating_fine_tuning} where $E_A, T_A, F_A$ are the audio encoder, temporal module and decoder of the pre-trained A2A-WaveGAN,
$E_V, T_V, F_V$ are the respective modules of V2A-WaveGAN, $D_V$ is the discriminator and $E_I$ is the identity encoder. We run two sets of experiments where $E_A, T_A$ are either fixed or they are fine-tuned with a learning rate of $1 \times 10^{-4}$, and we report the result with the lowest validation loss. All other modules are trained with a learning rate of $1 \times 10^{-4}$.
\end{itemize}

\noindent
In the experiments with fine-tuning we modify the batch normalization layers of the decoder by keeping track of separate running statistics depending on whether the temporal features were generated from the audio or video modalities. The batch normalization parameters are shared across modalities.

\subsection{Mel spectrogram model training and pre-training}
\label{subsection:mel_spectrogram_model_training_and_pretraining}

\subsubsection{A2A-MelSpec pre-training} We train using AdamW \cite{adamw} with a learning rate of $1 \times 10^{-3}, \beta_1 = 0.9, \beta_2 = 0.98$ and weight decay of $1 \times 10^{-2}$. We warm up the learning rate for 1 epoch and then decay it using a cosine schedule with warm restarts \cite{sgdr}, where $T_0 = 4, T_{mult} = 1$.

As with A2A-WaveGAN, we train in two steps: first, on all the audio datasets in Table \ref{table:audio_datasets} excluding CommonVoice until the validation loss is minimized and then we fine-tune the resulting best model on on all the audio datasets, including CommonVoice. In the second step, we reset the learning rate scheduler to start again from the warmup, while we continue with the learned model and optimizer parameters. For each version of A2A-MelSpec (VS, S, M) we select the model with the lowest validation loss.

\subsubsection{V2A-MelSpec training and finetuning} For all experiments we train using AdamW \cite{adamw} with $\beta_1 = 0.9, \beta_2 = 0.98$ and weight decay of $1 \times 10^{-2}$. We use a learning rate of $1 \times 10^{-3}$ for all experiments with seen speakers and $5 \times 10^{-4}$ for all experiments with unseen speakers. For experiments with GRID and TCD-TIMIT we warmup the learning rate for 20 epochs, while we do so for LRW for 15 epochs. Following the warmup, we decay the learning rate using a cosine schedule with warm restarts\cite{sgdr}, where $T_0 = 1, T_{mult} = 2$.

We conduct three sets of experiments with V2A-MelSpec:
\begin{itemize}
\item \textbf{From scratch:} Random initialization of all parameters
\item \textbf{Frozen pre-trained decoder:} Initializing the decoder with the pre-trained parameters of A2A-MelSpec and keeping it fixed during training.
\item \textbf{Fine-tuning pre-trained decoder:} Initializing the decoder with the pre-trained parameters of A2A-MelSpec and fine-tuning it with a lower learning rate.
\end{itemize}

\noindent
In the experiments using the pre-trained decoder, the learning rate for all other modules is the same as in the experiments from scratch. In addition, and for LRW only, we fine-tune the A2A-MelSpec model on the training set audio of LRW prior to conducting the last two sets of experiments, using a learning rate of $1 \times 10^{-4}$, and a 15 epoch warmup period followed by a cosine schedule. We found this step necessary to produce validation losses lower than training from scratch. Furthermore, and as in the experiments with raw waveforms, we kept track of separate running statistics in the batch normalization layers of the decoder: one set for temporal features generated from audio, and another for those generated from video inputs. The batch normalization parameters were also shared across modalities.

\subsection{Evaluation metrics}
We employ 4 objective metrics to measure the quality and intelligibility of our reconstructed speech: PESQ, STOI, ESTOI  and the word error rate (WER). Although it is widely acknowledged that no existing metrics correlate perfectly with human perception \cite{video_driven_speech_reconstruction_using_gans, end_to_end_video_to_speech_synthesis_using_gans}, these metrics are well established in the video-to-speech literature and are useful for comparison with other works.

PESQ (perceptual evaluation of speech quality)\cite{pesq} aims to capture the perceptual quality of the speech, originally created to measure speech quality of telephone networks and speech codecs. STOI (short-time objective intelligibility)\cite{stoi} and its extended version ESTOI \cite{estoi} aim to measure the intelligibility of speech samples. For all these metrics, higher scores are better.

WER measures the word-level accuracy of speech samples and we compute this using pre-trained speech recognition models, as is common practice. For experiments with GRID we use a pre-trained model \cite{visual_speech_recognition_for_multiple_languages_in_the_wild, svts} achieving a WER of 0.1\% on the real audio test set (based on the split in \cite{lipnet}). For LRW we use a model trained on LRW \cite{end_to_end_av_speech_recognition} and achieving a WER of 1.68\% on the test set. For TCD-TIMIT we did not employ WER as a metric as we were unable to find an accurate, publicly available speech recognition model for it.

We encourage the reader to visit our project website\footnotemark where we make available generated samples from our experiments.

\footnotetext{\texttt{\url{https://sites.google.com/view/v2a-with-audio-pretraining/home}}}

\section{Results}
In this section we present the results of our experiments, shown in Tables \ref{table:grid_4_seen_speakers} - \ref{table:lrw}. We use the datasets and splits defined in section \ref{subsection:datasets} and Table \ref{table:av_datasets}. In addition we split comparable methods into models that generate raw waveforms and those that generate acoustic features (such as mel spectrograms); we then compare our respective models within each category. To ensure a fair comparison we report results on the test set samples provided by the respective authors, except Lip2Wav\cite{lip2wav} where we report results as stated in the paper. For convenience, we append -F to our models using a face embedding and -SP when using a speaker embedding.

Our raw waveform models outperform previous work across all reconstruction metrics, irrespective of the identity embedding or of pre-training. They also outperform previous work on WER on GRID (33 speakers, seen) and LRW. This indicates that our proposed architecture and loss function result in a consistent improvement in generating waveforms from video.

In most cases, our mel spectrogram models also demonstrate improved metrics compared to previous work. We note that across all datasets, the fine-tuned V2A-MelSpec models result in a lower validation and test set loss compared to training from scratch. However, in experiments with GRID (33 speakers, seen) and LRW, V2A-MelSpec trained from scratch results in better reconstruction metrics and WER. This suggests that the loss function (L1 loss in this case) correlates imperfectly with the perceptual quality of the audio, in line with similar observations in the literature \cite{cmgan_interspeech}.

In addition, we observe that our models using a speaker embedding outperform those using a face embedding, which is expected given that the speaker embedding contains information about a speaker's voice.

\subsection{Results on seen speakers}

Tables \ref{table:grid_4_seen_speakers} - \ref{table:tcd_timit_lipspeakers} show our results on dataset splits involving seen speakers. For the GRID (4 speakers, seen) split (Table \ref{table:grid_4_seen_speakers}), V2A-WaveGAN-SP achieves the highest PESQ and ESTOI with with alternating fine-tuning, while the lowest WER is achieved using a face embedding and basic fine-tuning. We note that pre-training improves reconstruction metrics in V2A-WaveGAN, whereas WER is slightly more volatile. V2A-MelSpec-SP with fine-tuning shows improvement across all metrics compared to training from scratch. However, it is outpefrormed by VCA-GAN \cite{lip_to_speech_synthesis_with_visual_context_attention_gan} in PESQ and WER and by Lip2Wav\cite{lip2wav} in STOI and ESTOI.

% **************** GRID seen speakers (4 speakers) ******************** %
\begin{table}[h]
\captionsetup{justification=centering}
\caption{Results on GRID (4 speakers, seen)}
\begin{adjustbox}{width=\columnwidth}
\begin{tabular}{@{}lcccc@{}}
\toprule
\textbf{Method}                    & \textbf{PESQ$\uparrow$} & \textbf{STOI$\uparrow$} & \textbf{ESTOI$\uparrow$} & \textbf{WER (\%)$\downarrow$} \\ \midrule \midrule
\textbf{Raw waveform models} \\ \midrule \midrule
End-to-end WGAN (2018) \cite{video_driven_speech_reconstruction_using_gans}                           & 1.47             & 0.570             & 0.329              & 19.94                 \\
End-to-end WGAN (2022) \cite{end_to_end_video_to_speech_synthesis_using_gans}                            & 1.76             & 0.662             & 0.468              & \textbf{4.07}                 \\
\midrule
V2A-WaveGAN-F    & 1.82          & 0.681        & 0.492          & 5.50              \\
\hspace{3mm}+ basic fine-tuning     & 1.86          & 0.694         & 0.511               & 4.54                  \\
\hspace{3mm}+ alternating fine-tuning & 1.84          & 0.695         & 0.507          & 6.86              \\
\\
V2A-WaveGAN-SP & 1.87          & 0.693         & \textbf{0.513}        & 4.68              \\
\hspace{3mm}+ basic fine-tuning     & 1.87          & \textbf{0.695}       & 0.507          & 5.58              \\
\hspace{3mm}+ alternating fine-tuning & \textbf{1.90}          & 0.690         & \textbf{0.513}          & 4.99                       \\ \midrule \midrule

\textbf{Acoustic features models} \\ \midrule \midrule
Vid2Voc \cite{vocoder_based_speech_synthesis}                            & 1.61             & 0.650             & 0.455              & 9.29                 \\
Lip2Wav \cite{lip2wav}          & 1.77             & \textbf{0.731}             & \textbf{0.535}              & 14.08\footnotemark[2] \\
VCA-GAN \cite{lip_to_speech_synthesis_with_visual_context_attention_gan}                          & \textbf{2.03}             & 0.682             & 0.510  & \textbf{5.62} \\
Visual Voice Memory \cite{speech_reconstruction_visual_voice_memory}          & 1.82             & 0.643             & 0.481              & 6.08 \\
\midrule
V2A-MelSpec-VS-F    & 1.8          & 0.690        & 0.497          & 7.97              \\
\hspace{3mm}+ frozen pre-trained decoder     & 1.83          & 0.689         & 0.502               & 7.10                  \\
\hspace{3mm}+ fine-tuning pre-trained decoder & 1.82          & 0.690         & 0.502          & 6.70              \\
\\
V2A-MelSpec-VS-SP & 1.83          & 0.693         & 0.505        & 6.70              \\
\hspace{3mm}+ frozen pre-trained decoder     & 1.87          & 0.691       & 0.508          & 6.06              \\
\hspace{3mm}+ fine-tuning pre-trained decoder & 1.87          & 0.695         & 0.512          & 5.74              \\ \bottomrule
\end{tabular}
\end{adjustbox}
\vfill
\hspace{3mm} \\
$^2$Reported using Google speech-to-text API
\label{table:grid_4_seen_speakers}
\end{table}

% **************** GRID seen speakers (all speakers) ******************** %
\begin{table}[h]
\captionsetup{justification=centering}
\caption{Results on GRID (33 speakers, seen)}
\begin{adjustbox}{width=\columnwidth}
\begin{tabular}{@{}lcccc@{}}
\toprule
\textbf{Method}                    & \textbf{PESQ$\uparrow$} & \textbf{STOI$\uparrow$} & \textbf{ESTOI$\uparrow$} & \textbf{WER (\%)$\downarrow$} \\ \midrule \midrule
\textbf{Raw waveform models} \\ \midrule \midrule
End-to-end WGAN (2022) \cite{end_to_end_video_to_speech_synthesis_using_gans}                            & 1.70             & 0.667             & 0.465              & 4.59                 \\
\midrule
V2A-WaveGAN-F    & 1.94          & 0.707        & 0.512          & 3.60              \\
\hspace{3mm}+ basic fine-tuning     & 2.00          & 0.711         & 0.527               & 3.73                  \\
\hspace{3mm}+ alternating fine-tuning & 1.97          & 0.715         & 0.528          & 4.15              \\
\\
V2A-WaveGAN-SP & 2.00          & 0.712         & 0.529        & \textbf{2.79}              \\
\hspace{3mm}+ basic fine-tuning     & \textbf{2.07}          & \textbf{0.716}       & \textbf{0.539}          & 2.83              \\
\hspace{3mm}+ alternating fine-tuning & 2.01          & 0.715         & 0.532          & 3.52              \\ \midrule \midrule 

\textbf{Acoustic features models} \\ \midrule \midrule
VCA-GAN \cite{lip_to_speech_synthesis_with_visual_context_attention_gan}                            & 1.97             & 0.695             & 0.505              & 5.10                 \\
SVTS-S \cite{svts}                             & 1.97             & 0.705             & 0.523              & \textbf{2.37}                 \\
\midrule
V2A-MelSpec-S-F    & 1.96          & 0.715         & 0.529          & 3.08              \\
\hspace{3mm}+ frozen pre-trained decoder     & 1.94          & 0.709         & 0.518                & 4.21                   \\
\hspace{3mm}+ fine-tuning pre-trained decoder & 1.96          & 0.716         & 0.528          & 4.19              \\
\\
V2A-MelSpec-S-SP & \textbf{2.02}          & \textbf{0.720}         & 0.\textbf{538}          & 2.66              \\
\hspace{3mm}+ frozen pre-trained decoder     & 2.01          & 0.712         & 0.528          & 3.58              \\
\hspace{3mm}+ fine-tuning pre-trained decoder & 2.01          & 0.719         & 0.536          & 3.66              \\ \bottomrule
\end{tabular}
\end{adjustbox}
\label{table:grid_allspeakers_seen_speakers}
\end{table}

% ********************* TCD-TIMIT lipspeakers ************************* %
\begin{table}[h]
\captionsetup{justification=centering}
\caption{Results on TCD-TIMIT (3 lipspeakers, seen)}
\begin{adjustbox}{width=\columnwidth}
\begin{tabular}{@{}lcccc@{}}
\toprule
\textbf{Method}                   & \hspace{1cm} & \textbf{PESQ$\uparrow$} & \textbf{STOI$\uparrow$} & \textbf{ESTOI$\uparrow$} \\ \midrule \midrule
\textbf{Raw waveform models} \\ \midrule \midrule
End-to-end WGAN (2022) \cite{end_to_end_video_to_speech_synthesis_using_gans} &                           & 1.40             & 0.538             & 0.357                               \\
\midrule
V2A-WaveGAN-F &   & 1.39             & 0.543             & 0.362                               \\
\hspace{3mm}+ basic fine-tuning &    & \textbf{1.44}             & \textbf{0.568}             & \textbf{0.405}                               \\
\hspace{3mm}+ alternating fine-tuning & & 1.43             & 0.557             & 0.393                           \\
\\
V2A-WaveGAN-SP & & 1.41             & 0.552             & 0.364                               \\
\hspace{3mm}+ basic fine-tuning &    & 1.43             & 0.562             & 0.395                               \\
\hspace{3mm}+ alternating fine-tuning & & 1.43             & 0.565             & 0.402                             \\ \midrule \midrule
\textbf{Acoustic features models} \\ \midrule \midrule
VCA-GAN \cite{lip_to_speech_synthesis_with_visual_context_attention_gan} &                           & \textbf{1.43}             & \textbf{0.595}             & \textbf{0.420}  \\
Lip2Wav \cite{lip2wav} &                            & 1.35             & 0.558             & 0.365                               \\
\midrule
V2A-MelSpec-VS-F &   & 1.3          & 0.478         & 0.274                        \\
\hspace{3mm}+ frozen pre-trained decoder &    & 1.33          & 0.493         & 0.292                                   \\
\hspace{3mm}+ fine-tuning pre-trained decoder & & 1.35          & 0.491         & 0.305                        \\
\\
V2A-MelSpec-VS-SP & & 1.35          & 0.492         & 0.296                        \\
\hspace{3mm}+ frozen pre-trained decoder &     & 1.35          & 0.509         & 0.318                        \\
\hspace{3mm}+ fine-tuning pre-trained decoder & & 1.39          & 0.503         & 0.328                        \\ \bottomrule
\end{tabular}
\end{adjustbox}
\label{table:tcd_timit_lipspeakers}
\end{table}

With the GRID (33 speakers, seen) split (Table \ref{table:grid_allspeakers_seen_speakers}) V2A-WaveGAN-SP with basic fine-tuning outperforms all other raw waveform models across reconstruction metrics, but V2A-WaveGAN-SP trained from scratch achieves a slightly lower WER. V2A-MelSpec-SP trained from scratch also outperforms previous work across reconstruction metrics, but SVTS-S\cite{svts} shows a lower WER. Although fine-tuning the pre-trained decoder of V2A-MelSpec results in better metrics than using the frozen pre-trained decoder, the models trained from scratch report better metrics (except STOI in V2A-Melspec-F).

In experiments with TCD-TIMIT\cite{tcd_timit_databse} (3 lipspeakers, seen) V2A-WaveGAN-F with basic fine-tuning outperforms all other raw waveform models. In addition, both basic and alternating fine-tuning show a consistent improvement ovear training from scratch. Fine-tuning V2A-MelSpec models also results in improvement across all metrics compared to training from scratch. However VCA-GAN\cite{lip_to_speech_synthesis_with_visual_context_attention_gan} outperforms our V2A-MelSpec models on this dataset.

\subsection{Results on unseen speakers}

% ********************* GRID unseen speakers ************************* %
\begin{table}[h]
\captionsetup{justification=centering}
\caption{Results on GRID (33 speakers, unseen)}
\begin{adjustbox}{width=\columnwidth}
\begin{tabular}{@{}lcccc@{}}
\toprule
\textbf{Method}                    & \textbf{PESQ$\uparrow$} & \textbf{STOI$\uparrow$} & \textbf{ESTOI$\uparrow$} & \textbf{WER (\%)$\downarrow$} \\ \midrule \midrule
\textbf{Raw waveform models} \\ \midrule \midrule
End-to-end WGAN (2018) \cite{video_driven_speech_reconstruction_using_gans}                           & 1.26             & 0.494             & 0.198              & 32.76                 \\
End-to-end WGAN (2022) \cite{end_to_end_video_to_speech_synthesis_using_gans}                            & 1.37             & 0.568             & 0.289              & \textbf{16.05}                 \\
\midrule
V2A-WaveGAN-F    & 1.41          & 0.577        & 0.289          & 25.75              \\
\hspace{3mm}+ basic fine-tuning     & 1.42          & 0.593         & 0.316           & 18.57                  \\
\hspace{3mm}+ alternating fine-tuning & 1.41          & \textbf{0.596}         & 0.306          & 20.77              \\
\\
V2A-WaveGAN-SP & \textbf{1.43}          & 0.589         & 0.316        & 19.88              \\
\hspace{3mm}+ basic fine-tuning     & 1.41          & 0.595       & 0.325          & 17.08              \\
\hspace{3mm}+ alternating fine-tuning & \textbf{1.43}          & 0.595         & \textbf{0.326}          & 17.63              \\ \midrule \midrule
\textbf{Acoustic features models} \\ \midrule \midrule
Vid2Voc \cite{vocoder_based_speech_synthesis}   & 1.26             & 0.541             & 0.227              & 38.15 \\
VCA-GAN \cite{lip_to_speech_synthesis_with_visual_context_attention_gan}                            & 1.39             & 0.570             & 0.283              & 24.52                 \\
Visual Voice Memory \cite{speech_reconstruction_visual_voice_memory}          & 1.33             & 0.531             & 0.271              & 26.11                 \\
SVTS-S \cite{svts}                             & 1.40             & 0.588             & 0.318              & \textbf{17.84}                 \\
\midrule
V2A-MelSpec-S-F    & 1.35          & 0.577         & 0.298          & 23.25              \\
\hspace{3mm}+ frozen pre-trained decoder     & 1.38          & 0.584         & 0.297                & 22.86                   \\
\hspace{3mm}+ fine-tuning pre-trained decoder & 1.38          & 0.582         & 0.291          & 25.14              \\
\\
V2A-MelSpec-S-SP & 1.40          & 0.594         & 0.322          & 18.00              \\
\hspace{3mm}+ frozen pre-trained decoder     & 1.40         & 0.597         & 0.313          & 21.22              \\
\hspace{3mm}+ fine-tuning pre-trained decoder & \textbf{1.43}          & \textbf{0.598}         & \textbf{0.335}          & 17.90              \\ \bottomrule
\end{tabular}
\end{adjustbox}
\label{table:grid_unseen_speakers}
\end{table}

% ******************************* LRW *********************************** %
\begin{table}[h]
\captionsetup{justification=centering}
\caption{Results on LRW}
\begin{adjustbox}{width=\columnwidth}
\begin{tabular}{@{}lcccc@{}}
\toprule
\textbf{Method}                    & \textbf{PESQ$\uparrow$} & \textbf{STOI$\uparrow$} & \textbf{ESTOI$\uparrow$} & \textbf{WER (\%)$\downarrow$} \\ \midrule \midrule
\textbf{Raw waveform models} \\ \midrule \midrule
End-to-end WGAN (2022) \cite{end_to_end_video_to_speech_synthesis_using_gans}                            & 1.33             & 0.552             & 0.331              & 42.38                 \\
\midrule
V2A-WaveGAN-F    & 1.39          & 0.590        & 0.378          & 41.00              \\
\hspace{3mm}+ basic fine-tuning     & 1.41          & 0.603         & 0.395           & 35.83                  \\
\hspace{3mm}+ alternating fine-tuning & 1.41          & 0.606         & 0.402          & 35.38              \\
\\
V2A-WaveGAN-SP & 1.46          & 0.623         & 0.445        & \textbf{29.79}              \\
\hspace{3mm}+ basic fine-tuning     & 1.47          & 0.630       & 0.443          & 29.88              \\
\hspace{3mm}+ alternating fine-tuning & \textbf{1.48}          & \textbf{0.637}         & \textbf{0.456}          & 30.00              \\ \midrule \midrule
\textbf{Acoustic features models} \\ \midrule \midrule
VCA-GAN \cite{lip_to_speech_synthesis_with_visual_context_attention_gan}                            & 1.34             & 0.565             & 0.364              & 37.07                 \\
Lip2Wav \cite{lip2wav}          & 1.20             & 0.543             & 0.344              & 34.20\footnotemark[2]                 \\
SVTS-M \cite{svts}                             & 1.46             & \textbf{0.649}             & 0.482              & \textbf{12.90}                 \\
\midrule
V2A-MelSpec-M-F    & 1.40          & 0.612         & 0.426          & 22.67              \\
\hspace{3mm}+ frozen pre-trained decoder     & 1.38          & 0.596         & 0.399                & 34.44                   \\
\hspace{3mm}+ fine-tuning pre-trained decoder & 1.39          & 0.605         & 0.412          & 28.84              \\
\\
V2A-MelSpec-M-SP & \textbf{1.48}          & \textbf{0.649}         & \textbf{0.484}          & 14.96              \\
\hspace{3mm}+ frozen pre-trained decoder     & 1.45         & 0.633         & 0.455          & 23.88              \\
\hspace{3mm}+ fine-tuning pre-trained decoder & 1.46          & 0.646         & 0.476          & 18.70              \\ \bottomrule
\end{tabular}
\end{adjustbox}
\vfill
\hspace{3mm} \\
$^2$Reported using Google speech-to-text API
\label{table:lrw}
\end{table}

Tables \ref{table:grid_unseen_speakers} and \ref{table:lrw} show our results on dataset splits involving unseen speakers. On GRID (33 speakers, unseen) (Table \ref{table:grid_unseen_speakers}) V2A-WaveGAN-F with alternating fine-tuning reports the best STOI and V2A-WaveGAN-SP with alternating fine-tuning reports the best PESQ and ESTOI. V2A-MelSpec-S-SP with fine-tuning outperforms previous works on the reconstruction metrics but SVTS-S \cite{svts} shows a slightly lower WER. 

Finally, we report our results on LRW in Table \ref{table:lrw}. V2A-WaveGAN-SP with alternating fine-tuning outperforms all other raw waveform models on reconstruction metrics, while this model trained from scratch shows the lowest WER. V2A-MelSpec-M-SP trained from scratch outperforms previous work on reconstruction metrics (equalling SVTS-M on STOI) but SVTS-M\cite{svts} reports a lower WER. As with GRID all speakers (seen), our V2A-MelSpec models trained from scratch report better metrics than those with pre-training despite the latter exhibiting lower validation and test set losses. Similarly, fine-tuning the pre-trained decoder produces audio of higher quality than using a frozen pre-trained decoder.

\section{Conclusion}
In this paper we have investigated the pre-training of encoder-decoder audio models on a large volume of speech data and then initializing the decoder of video-to-speech synthesis models using the pre-trained learned parameters. We have proposed generative models in the raw waveform and mel spectrogram domains, pre-training approaches using speech data as well as methodologies to fine-tune the video-to-speech models using the pre-trained audio-to-audio models.

We believe it would be beneficial to extend these pre-training approaches to other audio-visual speech tasks, such as speech enhancement and inpainting. It would also be useful to experiment with different temporal modules and downsampling-upsampling schemes for the audio-to-audio models. Furthermore, it would be interesting to examine further the settings where loading the state of the decoder optimizer during initialization, in addition to the pre-trained decoder parameters, improves the generated audio. Another promising research direction would be to investigate alternative pre-training tasks relying on self-supervision, such as speech denoising and the use of masked autoencoders.

\bibliographystyle{IEEEtran}
\bibliography{references}

\clearpage
\section{Supplementary Material}
\beginsupplement

\subsection{Fine-tuning configuration}

Tables \ref{supplementary:table:finetuning_configuration_v2a_wavegan} and \ref{supplementary:table:finetuning_configuration_v2a_melspec} outline the fine-tuning configuration for the V2A-WaveGAN and V2A-MelSpec models.

\begin{table*}
\centering
\captionsetup{justification=centering}
\caption{Fine-tuning configuration of V2A-WaveGAN}
\begin{tabular}{@{}llccc@{}}
\toprule
\textbf{Dataset}        & \textbf{Fine-tuning method} & \textbf{Load Decoder optimizer?} & \textbf{Load Discriminator optimizer?} & \textbf{Audio encoder learning rate}               \\ \midrule
GRID (4 speakers, seen)    & Basic (F, SP)               & No                               & No                                     & -                                                  \\
                        & Alternating (F, SP)         & -                                & -                                      & 0                                                  \\ \midrule
GRID (33 speakers, seen)  & Basic (F, SP)               & Yes                              & Yes                                    & -                                                  \\
                        & Alternating (F, SP)         & -                                & -                                      & $1 \times 10^{-4}$ \\ \midrule
GRID (33 speakers, unseen)    & Basic (F, SP)               & No                               & No                                     & -                                                  \\
                        & Alternating (F)             & -                                & -                                      & $1 \times 10^{-4}$ \\
                        & Alternating (SP)            & -                                & -                                      & 0                                                  \\ \midrule
TCD-TIMIT (3 lipspeakers, seen) & Basic (F, SP)               & No                               & No                                     & -                                                  \\
                        & Alternating (F)             & -                                & -                                      & 0                                                  \\
                        & Alternating (S)             & -                                & -                                      & $1 \times 10^{-4}$ \\ \midrule
LRW                     & Basic (F, SP)                      & Yes                              & Yes                                    & -                                                  \\
                        & Alternating (F)             & -                                & -                                      & $1 \times 10^{-4}$ \\
                        & Alternating (SP)             & -                                & -                                      & 0                                                  \\ \bottomrule
\end{tabular}
\vfill
\hspace{3mm} \\
F = Face embedding, SP = Speaker embedding
\label{supplementary:table:finetuning_configuration_v2a_wavegan}
\end{table*}

\begin{table*}
\centering
\captionsetup{justification=centering}
\caption{Fine-tuning configuration of V2A-MelSpec}
\begin{tabular}{@{}llccc@{}}
\toprule
\textbf{Dataset}        & \textbf{Fine-tuning method}             & \textbf{\begin{tabular}[c]{@{}c@{}}\# of epochs of\\ frozen Decoder\end{tabular}} & \textbf{Load Decoder optimizer?} & \textbf{Decoder learning rate} \\ \midrule
All    & Frozen pre-trained decoder (F, SP) & All                                      & -                               & -                       \\
    &  &                                       &                               &                       \\ \midrule
GRID (4 speakers, seen)    & Fine-tuning pre-trained decoder (F, SP) & 20                                      & No                               & $1 \times 10^{-4}$                       \\ \midrule
GRID (33 speakers, seen)  & Fine-tuning pre-trained decoder (F)     & 20                                      & No                               & $1 \times 10^{-4}$                       \\
                        & Fine-tuning pre-trained decoder (SP)    & 20                                      & Yes                              & $1 \times 10^{-4}$                       \\ \midrule
GRID (33 speakers, unseen)    & Fine-tuning pre-trained decoder (F, SP) & 0                                       & No                               & $1 \times 10^{-5}$                       \\ \midrule
TCD-TIMIT (3 lipspeakers, seen) & Fine-tuning pre-trained decoder (F, SP) & 20                                      & No                               & $1 \times 10^{-7}$                       \\ \midrule
LRW                     & Fine-tuning pre-trained decoder (F, SP) & 0                                       & Yes                              & $1 \times 10^{-5}$                       \\ \bottomrule
\end{tabular}
\vfill
\hspace{3mm} \\
F = Face embedding, SP = Speaker embedding
\label{supplementary:table:finetuning_configuration_v2a_melspec}
\end{table*}

\subsection{Loading the states of the pre-trained optimizers}

In sections \ref{subsection:raw_waveform_model_training_and_pretraining} and \ref{subsection:mel_spectrogram_model_training_and_pretraining} we described the fine-tuning procedures, involving initialization of the decoder (and discriminator for raw waveforms) and optionally initializing the respective optimizer(s) with the states of the optimizer(s) of the pre-trained A2A model. The decision to initialize the optimizers as such was made by reference to the validation loss during training. We will mention two examples to illustrate this.

\begin{figure}[h]
  \includegraphics[width=\linewidth]{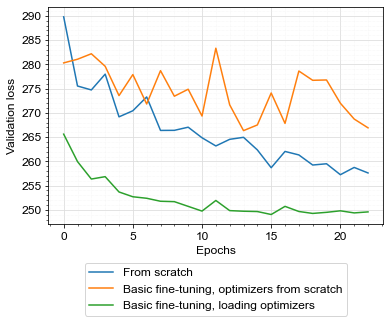}
\caption{Training V2A-WaveGAN-F on LRW}
\label{lrw_optimizer}
\end{figure}

\begin{figure}[h]
  \includegraphics[width=\linewidth]{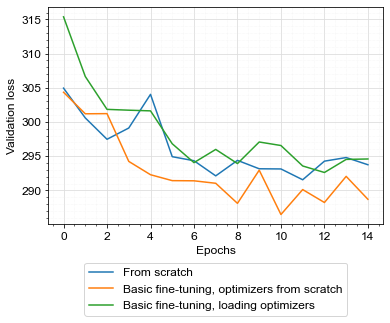}
\caption{Training V2A-WaveGAN-SP on GRID (33 speakers, unseen)}
\label{grid_unseen_optimizer}
\end{figure}

Fig. \ref{lrw_optimizer} illustrates the validation loss when training V2A-WaveGAN-F on LRW. If we fine-tune by loading the pre-trained decoder and discriminator, and initialize their optimizers using their default initialization we obtain a higher validation loss than by training the model from scratch. By loading the states of the decoder and discriminator optimizers as well we see that fine-tuning progresses with consistently lower validation loss.

However, we observe the opposite when training V2A-WaveGAN-SP on the GRID (33 speakers, unseen) split (Fig. \ref{grid_unseen_optimizer}). In this case we obtain a lower validation loss by initializing the optimizers with their default initialization, than by loading their pre-trained states. We note that in both examples above, the wrong optimizer initialization decision leads to higher validation loss than training from scratch. 

\subsection{Fine-tuning and batch normalization statistics}

\begin{figure}[h]
  \includegraphics[width=\linewidth]{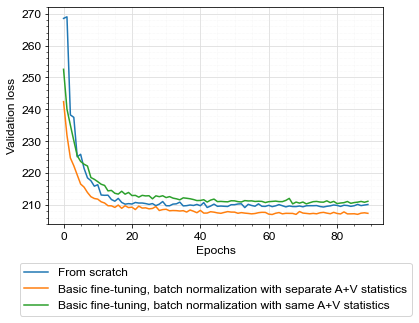}
\caption{Training V2A-WaveGAN-SP on GRID (33 speakers, seen)}
\label{grid_allspeakers_seen_batchnorm}
\end{figure}

In sections \ref{subsection:raw_waveform_model_training_and_pretraining} and \ref{subsection:mel_spectrogram_model_training_and_pretraining}, where we described the fine-tuning procedures, we explained that we kept track of separate statistics for temporal features (inputs to the decoder) generated from audio and those generated from video inputs. We demonstrate the importance of this step in Fig. \ref{grid_allspeakers_seen_batchnorm}. By fine-tuning with the same audio and video batch normalization statistics (i.e. by updating the statistics computed during the A2A model pre-training) we observe a higher validation loss compared to training from scratch. When we keep separate running statistics, i.e. begin a new set of batch normalization statistics for video-to-audio generation, the ensuing validation loss is consistently lower than training from scratch.

\subsection{Perceptual loss for raw waveform models}

\subsubsection{V2A-WaveGAN}
While conducting the sequential tuning experiments in Section \ref{subsubsection:raw_waveform_generation_loss_function}, we noticed a tradeoff: as we increased the values of the reconstruction loss coefficients (and thus the relative importance of reconstruction accuracy vs. learning the reference distribution) the synthesized speech became clearer but with lower realism. Comparing our found coefficients to those of loss functions in relevant vocoder GANs (e.g. \cite{parallel_wavegan, multi_band_melgan}) we note that the latter place relatively higher importance on the adversarial loss. Therefore, we construct an additional generator loss function with a second set of coefficients where $\lambda_1 = 2.5$, $\lambda_2 = 1.0$ as in \cite{multi_band_melgan} and $\lambda_3 = 0.1$. The value of $\lambda_3$ was set by manual inspection, such that the contribution of the MFCC loss to the total loss followed the contribution of the multi-resolution STFT loss during training. We did not arrive at these coefficient values through any further optimization. We denote the model trained with these as V2A-WaveGAN-P, where P stands for perceptual.

\subsubsection{A2A-WaveGAN}
Similarly, we train an additional A2A model with these loss function coefficients, denoted by A2A-WaveGAN-P. \\

\noindent
Fig. \ref{supplementary:table:grid_4_seen_speakers} - \ref{supplementary:table:lrw} show the results of our experiments, including V2A-WaveGAN-P.

% **************** GRID seen speakers (4 speakers) ******************** %
\begin{table}[h]
\captionsetup{justification=centering}
\caption{Results on GRID (4 speakers, seen)}
\begin{adjustbox}{width=\columnwidth}
\begin{tabular}{@{}lcccc@{}}
\toprule
\textbf{Method}                    & \textbf{PESQ$\uparrow$} & \textbf{STOI$\uparrow$} & \textbf{ESTOI$\uparrow$} & \textbf{WER (\%)$\downarrow$} \\ \midrule \midrule
\textbf{Raw waveform models} \\ \midrule \midrule
End-to-end WGAN (2018) \cite{video_driven_speech_reconstruction_using_gans}                           & 1.47             & 0.570             & 0.329              & 19.94                 \\
End-to-end WGAN (2022) \cite{end_to_end_video_to_speech_synthesis_using_gans}                            & 1.76             & 0.662             & 0.468              & \textbf{4.07}                 \\
\midrule
V2A-WaveGAN-F    & 1.82          & 0.681        & 0.492          & 5.50              \\
\hspace{3mm}+ basic fine-tuning     & 1.86          & 0.694         & 0.511               & 4.54                  \\
\hspace{3mm}+ alternating fine-tuning & 1.84          & 0.695         & 0.507          & 6.86              \\
\\
V2A-WaveGAN-SP & 1.87          & 0.693         & \textbf{0.513}        & 4.68              \\
\hspace{3mm}+ basic fine-tuning     & 1.87          & \textbf{0.695}       & 0.507          & 5.58              \\
\hspace{3mm}+ alternating fine-tuning & \textbf{1.90}          & 0.690         & \textbf{0.513}          & 4.99                       \\ \midrule
V2A-WaveGAN-P-F    & 1.70          & 0.674        & 0.488          & 6.22              \\
\hspace{3mm}+ basic fine-tuning     & 1.79          & 0.675         & 0.497               & 5.18                  \\
\hspace{3mm}+ alternating fine-tuning & 1.81          & 0.680         & 0.507          & 4.37                          \\
\\
V2A-WaveGAN-P-SP & 1.69          & 0.667         & 0.467        & 8.33              \\
\hspace{3mm}+ basic fine-tuning     & 1.83          & 0.677       & 0.504          & 4.22              \\
\hspace{3mm}+ alternating fine-tuning & 1.85          & 0.681         & 0.510          & 4.45                          \\ \midrule \midrule

\textbf{Acoustic features models} \\ \midrule \midrule
Vid2Voc \cite{vocoder_based_speech_synthesis}                            & 1.61             & 0.650             & 0.455              & 9.29                 \\
Lip2Wav \cite{lip2wav}          & 1.77             & \textbf{0.731}             & \textbf{0.535}              & 14.08\footnotemark[1] \\
VCA-GAN \cite{lip_to_speech_synthesis_with_visual_context_attention_gan}                          & \textbf{2.03}             & 0.682             & 0.510  & \textbf{5.62} \\
Visual Voice Memory \cite{speech_reconstruction_visual_voice_memory}          & 1.82             & 0.643             & 0.481              & 6.08 \\
\midrule
V2A-MelSpec-VS-F    & 1.8          & 0.690        & 0.497          & 7.97              \\
\hspace{3mm}+ frozen pre-trained decoder     & 1.83          & 0.689         & 0.502               & 7.10                  \\
\hspace{3mm}+ fine-tuning pre-trained decoder & 1.82          & 0.690         & 0.502          & 6.70              \\
\\
V2A-MelSpec-VS-SP & 1.83          & 0.693         & 0.505        & 6.70              \\
\hspace{3mm}+ frozen pre-trained decoder     & 1.87          & 0.691       & 0.508          & 6.06              \\
\hspace{3mm}+ fine-tuning pre-trained decoder & 1.87          & 0.695         & 0.512          & 5.74              \\ \bottomrule
\end{tabular}
\end{adjustbox}
\vfill
\hspace{3mm} \\
$^1$Reported using Google speech-to-text API
\label{supplementary:table:grid_4_seen_speakers}
\end{table}

% **************** GRID seen speakers (all speakers) ******************** %
\begin{table}
\captionsetup{justification=centering}
\caption{Results on GRID (33 speakers, seen)}
\begin{adjustbox}{width=\columnwidth}
\begin{tabular}{@{}lcccc@{}}
\toprule
\textbf{Method}                    & \textbf{PESQ$\uparrow$} & \textbf{STOI$\uparrow$} & \textbf{ESTOI$\uparrow$} & \textbf{WER (\%)$\downarrow$} \\ \midrule \midrule
\textbf{Raw waveform models} \\ \midrule \midrule
End-to-end WGAN (2022) \cite{end_to_end_video_to_speech_synthesis_using_gans}                            & 1.70             & 0.667             & 0.465              & 4.59                 \\
\midrule
V2A-WaveGAN-F    & 1.94          & 0.707        & 0.512          & 3.60              \\
\hspace{3mm}+ basic fine-tuning     & 2.00          & 0.711         & 0.527               & 3.73                  \\
\hspace{3mm}+ alternating fine-tuning & 1.97          & 0.715         & 0.528          & 4.15              \\
\\
V2A-WaveGAN-SP & 2.00          & 0.712         & 0.529        & \textbf{2.79}              \\
\hspace{3mm}+ basic fine-tuning     & \textbf{2.07}          & \textbf{0.716}       & \textbf{0.539}          & 2.83              \\
\hspace{3mm}+ alternating fine-tuning & 2.01          & 0.715         & 0.532          & 3.52              \\
\midrule
V2A-WaveGAN-P-F    & 1.82          & 0.680        & 0.495          & 3.28              \\
\hspace{3mm}+ basic fine-tuning     & 1.88          & 0.688         & 0.505               & 3.87                  \\
\hspace{3mm}+ alternating fine-tuning & 1.90          & 0.690         & 0.509          & 3.69              \\
\\
V2A-WaveGAN-P-SP & 1.88          & 0.689         & 0.508        & 2.86              \\
\hspace{3mm}+ basic fine-tuning     & 1.93          & 0.695       & 0.515          & 3.83              \\
\hspace{3mm}+ alternating fine-tuning & 1.97          & 0.701         & 0.523          & 3.30              \\ \midrule \midrule 

\textbf{Acoustic features models} \\ \midrule \midrule
VCA-GAN \cite{lip_to_speech_synthesis_with_visual_context_attention_gan}                            & 1.97             & 0.695             & 0.505              & 5.10                 \\
SVTS-S \cite{svts}                             & 1.97             & 0.705             & 0.523              & \textbf{2.37}                 \\
\midrule
V2A-MelSpec-S-F    & 1.96          & 0.715         & 0.529          & 3.08              \\
\hspace{3mm}+ frozen pre-trained decoder     & 1.94          & 0.709         & 0.518                & 4.21                   \\
\hspace{3mm}+ fine-tuning pre-trained decoder & 1.96          & 0.716         & 0.528          & 4.19              \\
\\
V2A-MelSpec-S-SP & \textbf{2.02}          & \textbf{0.720}         & 0.\textbf{538}          & 2.66              \\
\hspace{3mm}+ frozen pre-trained decoder     & 2.01          & 0.712         & 0.528          & 3.58              \\
\hspace{3mm}+ fine-tuning pre-trained decoder & 2.01          & 0.719         & 0.536          & 3.66              \\ \bottomrule
\end{tabular}
\end{adjustbox}
\label{supplementary:table:grid_allspeakers_seen_speakers}
\end{table}
% ************************************************************************ %

% ********************* TCD-TIMIT lipspeakers ************************* %
\begin{table}[h]
\captionsetup{justification=centering}
\caption{Results on TCD-TIMIT (3 lipspeakers, seen)}
\begin{adjustbox}{width=\columnwidth}
\begin{tabular}{@{}lcccc@{}}
\toprule
\textbf{Method}                   & \hspace{1cm} & \textbf{PESQ$\uparrow$} & \textbf{STOI$\uparrow$} & \textbf{ESTOI$\uparrow$} \\ \midrule \midrule
\textbf{Raw waveform models} \\ \midrule \midrule
End-to-end WGAN (2022) \cite{end_to_end_video_to_speech_synthesis_using_gans} &                           & 1.40             & 0.538             & 0.357                               \\
\midrule
V2A-WaveGAN-F &   & 1.39             & 0.543             & 0.362                               \\
\hspace{3mm}+ basic fine-tuning &    & \textbf{1.44}             & \textbf{0.568}             & \textbf{0.405}                               \\
\hspace{3mm}+ alternating fine-tuning & & 1.43             & 0.557             & 0.393                           \\
\\
V2A-WaveGAN-SP & & 1.41             & 0.552             & 0.364                               \\
\hspace{3mm}+ basic fine-tuning &    & 1.43             & 0.562             & 0.395                               \\
\hspace{3mm}+ alternating fine-tuning & & 1.43             & 0.565             & 0.402                             \\ 
\midrule
V2A-WaveGAN-P-F    & & 1.27             & 0.465             & 0.256                             \\
\hspace{3mm}+ basic fine-tuning     & & 1.39             & 0.530             & 0.361                               \\
\hspace{3mm}+ alternating fine-tuning & & 1.39             & 0.551             & 0.397                               \\
\\
V2A-WaveGAN-P-SP & & 1.34             & 0.490             & 0.394                             \\
\hspace{3mm}+ basic fine-tuning     & & 1.38             & 0.523             & 0.344                              \\
\hspace{3mm}+ alternating fine-tuning & & 1.41             & 0.556             & 0.401                              \\ \midrule \midrule

\textbf{Acoustic features models} \\ \midrule \midrule
VCA-GAN \cite{lip_to_speech_synthesis_with_visual_context_attention_gan} &                           & \textbf{1.43}             & \textbf{0.595}             & \textbf{0.420}  \\
Lip2Wav \cite{lip2wav} &                            & 1.35             & 0.558             & 0.365                               \\
\midrule
V2A-MelSpec-VS-F &   & 1.3          & 0.478         & 0.274                        \\
\hspace{3mm}+ frozen pre-trained decoder &    & 1.33          & 0.493         & 0.292                                   \\
\hspace{3mm}+ fine-tuning pre-trained decoder & & 1.35          & 0.491         & 0.305                        \\
\\
V2A-MelSpec-VS-SP & & 1.35          & 0.492         & 0.296                        \\
\hspace{3mm}+ frozen pre-trained decoder &     & 1.35          & 0.509         & 0.318                        \\
\hspace{3mm}+ fine-tuning pre-trained decoder & & 1.39          & 0.503         & 0.328                        \\ \bottomrule
\end{tabular}
\end{adjustbox}
\label{supplementary:table:tcd_timit_lipspeakers}
\end{table} 
% ************************************************************************ %

% ********************* GRID unseen speakers ************************* %
\begin{table}[h]
\captionsetup{justification=centering}
\caption{Results on GRID (33 speakers, unseen)}
\begin{adjustbox}{width=\columnwidth}
\begin{tabular}{@{}lcccc@{}}
\toprule
\textbf{Method}                    & \textbf{PESQ$\uparrow$} & \textbf{STOI$\uparrow$} & \textbf{ESTOI$\uparrow$} & \textbf{WER (\%)$\downarrow$} \\ \midrule \midrule
\textbf{Raw waveform models} \\ \midrule \midrule
End-to-end WGAN (2018) \cite{video_driven_speech_reconstruction_using_gans}                           & 1.26             & 0.494             & 0.198              & 32.76                 \\
End-to-end WGAN (2022) \cite{end_to_end_video_to_speech_synthesis_using_gans}                            & 1.37             & 0.568             & 0.289              & \textbf{16.05}                 \\
\midrule
V2A-WaveGAN-F    & 1.41          & 0.577        & 0.289          & 25.75              \\
\hspace{3mm}+ basic fine-tuning     & 1.42          & 0.593         & 0.316           & 18.57                  \\
\hspace{3mm}+ alternating fine-tuning & 1.41          & \textbf{0.596}         & 0.306          & 20.77              \\
\\
V2A-WaveGAN-SP & \textbf{1.43}          & 0.589         & 0.316        & 19.88              \\
\hspace{3mm}+ basic fine-tuning     & 1.41          & 0.595       & 0.325          & 17.08              \\
\hspace{3mm}+ alternating fine-tuning & \textbf{1.43}          & 0.595         & \textbf{0.326}          & 17.63              \\ 
\midrule
V2A-WaveGAN-P-F    & 1.32          & 0.549        & 0.293          & 18.43              \\
\hspace{3mm}+ basic fine-tuning     & 1.33          & 0.558         & 0.295               & 18.74                  \\
\hspace{3mm}+ alternating fine-tuning & 1.37          & 0.574         & 0.303          & 18.03              \\
\\
V2A-WaveGAN-P-SP & 1.33          & 0.558         & 0.287        & 21.72              \\
\hspace{3mm}+ basic fine-tuning     & 1.34          & 0.571       & 0.305          & 17.95              \\
\hspace{3mm}+ alternating fine-tuning & 1.37          & 0.576         & 0.317          & 17.26             \\ \midrule \midrule
\textbf{Acoustic features models} \\ \midrule \midrule
Vid2Voc \cite{vocoder_based_speech_synthesis}   & 1.26             & 0.541             & 0.227              & 38.15 \\
VCA-GAN \cite{lip_to_speech_synthesis_with_visual_context_attention_gan}                            & 1.39             & 0.570             & 0.283              & 24.52                 \\
Visual Voice Memory \cite{speech_reconstruction_visual_voice_memory}          & 1.33             & 0.531             & 0.271              & 26.11                 \\
SVTS-S \cite{svts}                             & 1.40             & 0.588             & 0.318              & \textbf{17.84}                 \\
\midrule
V2A-MelSpec-S-F    & 1.35          & 0.577         & 0.298          & 23.25              \\
\hspace{3mm}+ frozen pre-trained decoder     & 1.38          & 0.584         & 0.297                & 22.86                   \\
\hspace{3mm}+ fine-tuning pre-trained decoder & 1.38          & 0.582         & 0.291          & 25.14              \\
\\
V2A-MelSpec-S-SP & 1.40          & 0.594         & 0.322          & 18.00              \\
\hspace{3mm}+ frozen pre-trained decoder     & 1.40         & 0.597         & 0.313          & 21.22              \\
\hspace{3mm}+ fine-tuning pre-trained decoder & \textbf{1.43}          & \textbf{0.598}         & \textbf{0.335}          & 17.90              \\ \bottomrule
\end{tabular}
\end{adjustbox}
\label{supplementary:table:grid_unseen_speakers}
\end{table}
% *********************************************************************** %

% ******************************* LRW *********************************** %
\begin{table}[h]
\captionsetup{justification=centering}
\caption{Results on LRW}
\begin{adjustbox}{width=\columnwidth}
\begin{tabular}{@{}lcccc@{}}
\toprule
\textbf{Method}                    & \textbf{PESQ$\uparrow$} & \textbf{STOI$\uparrow$} & \textbf{ESTOI$\uparrow$} & \textbf{WER (\%)$\downarrow$} \\ \midrule \midrule
\textbf{Raw waveform models} \\ \midrule \midrule
End-to-end WGAN (2022) \cite{end_to_end_video_to_speech_synthesis_using_gans}                            & 1.33             & 0.552             & 0.331              & 42.38                 \\
\midrule
V2A-WaveGAN-F    & 1.39          & 0.590        & 0.378          & 41.00              \\
\hspace{3mm}+ basic fine-tuning     & 1.41          & 0.603         & 0.395           & 35.83                  \\
\hspace{3mm}+ alternating fine-tuning & 1.41          & 0.606         & 0.402          & 35.38              \\
\\
V2A-WaveGAN-SP & 1.46          & 0.623         & 0.445        & \textbf{29.79}              \\
\hspace{3mm}+ basic fine-tuning     & 1.47          & 0.630       & 0.443          & 29.88              \\
\hspace{3mm}+ alternating fine-tuning & \textbf{1.48}          & \textbf{0.637}         & \textbf{0.456}          & 30.00              \\
\midrule
V2A-WaveGAN-P-F    & 1.22          & 0.467        & 0.238          & 66.62              \\
\hspace{3mm}+ basic fine-tuning     & 1.23          & 0.499         & 0.282               & 55.82                  \\
\hspace{3mm}+ alternating fine-tuning & 1.24          & 0.503         & 0.295          & 47.57              \\
\\
V2A-WaveGAN-P-SP & 1.22          & 0.482         & 0.275        & 61.84              \\
\hspace{3mm}+ basic fine-tuning     & 1.30          & 0.568       & 0.374          & 34.21              \\
\hspace{3mm}+ alternating fine-tuning & 1.30          & 0.560         & 0.367          & 33.67              \\ \midrule \midrule
\textbf{Acoustic features models} \\ \midrule \midrule
VCA-GAN \cite{lip_to_speech_synthesis_with_visual_context_attention_gan}                            & 1.34             & 0.565             & 0.364              & 37.07                 \\
Lip2Wav \cite{lip2wav}          & 1.20             & 0.543             & 0.344              & 34.20\footnotemark[2]                 \\
SVTS-M \cite{svts}                             & 1.46             & \textbf{0.649}             & 0.482              & \textbf{12.90}                 \\
\midrule
V2A-MelSpec-M-F    & 1.40          & 0.612         & 0.426          & 22.67              \\
\hspace{3mm}+ frozen pre-trained decoder     & 1.38          & 0.596         & 0.399                & 34.44                   \\
\hspace{3mm}+ fine-tuning pre-trained decoder & 1.39          & 0.605         & 0.412          & 28.84              \\
\\
V2A-MelSpec-M-SP & \textbf{1.48}          & \textbf{0.649}         & \textbf{0.484}          & 14.96              \\
\hspace{3mm}+ frozen pre-trained decoder     & 1.45         & 0.633         & 0.455          & 23.88              \\
\hspace{3mm}+ fine-tuning pre-trained decoder & 1.46          & 0.646         & 0.476          & 18.70              \\ \bottomrule
\end{tabular}
\end{adjustbox}
\vfill
\hspace{3mm} \\
$^1$Reported using Google speech-to-text API
\label{supplementary:table:lrw}
\end{table}

%%%%%%%%%%%%%%%%%%%%%%%%%%%%%%%%%%%%%%%%%%%%%%%%%%%%%%%%%%%%%%%%%%%%%%%%%%%%%%
\clearpage
\subsection{Architectural details - Video frames encoder}

\begin{figure}[h]
  \includegraphics[width=18cm]{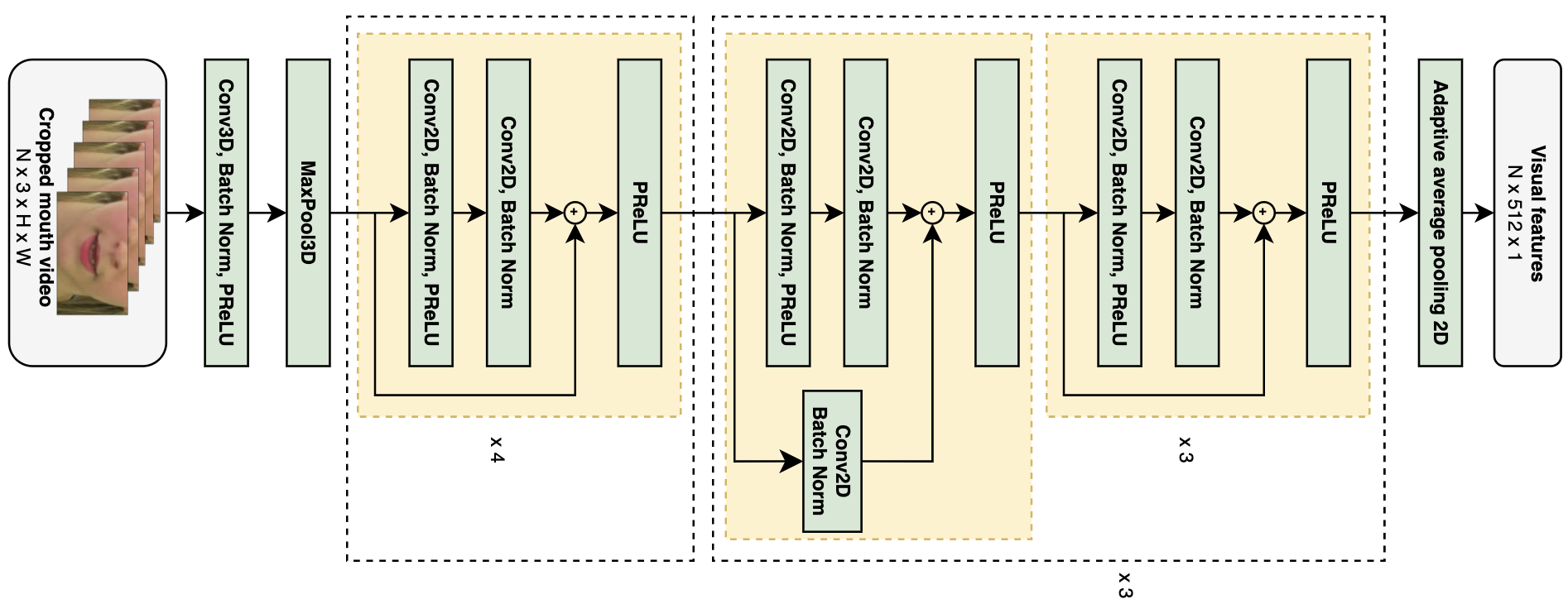}
\caption{Video frames encoder}
\label{video_frames_encoder}
\end{figure}

\begin{table}[h]
\caption{Architecture of a residual stack of the video frames encoder with $C$ channels and stride $S$}
\begin{tabular}{@{}lcccc@{}}
\toprule
\textbf{Layer}       & \textbf{Channels} & \textbf{Kernel} & \textbf{Stride} & \textbf{Padding} \\ \midrule
Conv 2D + BatchNorm 2D + PReLU & $C$                 & (3, 3)          & ($S$, $S$)          & (1, 1)           \\
Conv 2D + BatchNorm 2D         & $C$               & (3, 3)          & (1, 1)          & (1, 1)           \\
      &                   &                 &                 &                  \\
\textbf{Skip layer}       &                   &                 &                 &                  \\
\textbf{if downsample:}       &                   &                 &                 &                  \\
\hspace{3mm}Conv 2D + BatchNorm 2D         & $C$               & (1, 1)          & ($S$, $S$)          & -                \\
\textbf{else:}       &                   &                 &                 &                  \\
\hspace{3mm}Identity         & -               & -          & -          & -                \\ \bottomrule
\end{tabular}
\end{table}

\begin{table}[h]
\captionsetup{justification=centering}
\caption{Architecture of video frames encoder}
\begin{tabular}{@{}lccccc@{}}
\toprule
\textbf{Layer}           & \textbf{Channels} & \textbf{Kernel} & \textbf{Stride} & \textbf{Dilation} & \textbf{Padding} \\ \midrule
Conv 3D + BatchNorm 3D + PReLU     & 64                & (5, 7, 7)       & (1, 2, 2)       & 1                 & (2, 3, 3)        \\
MaxPool 3D               &                   & (1, 3, 3)       & (1, 2, 2)       & 1                 & (0, 1, 1)        \\
Residual Stack + PreLU x 4          & 64                &                 & (1, 1)          &                   &                  \\
Residual Stack (downsample) + PreLU & 128               &                 & (2, 2)          &                   &                  \\
Residual Stack + PreLU x 3          & 128               &                 & (1, 1)          &                   &                  \\
Residual Stack (downsample) + PreLU & 256               &                 & (2, 2)          &                   &                  \\
Residual Stack + PreLU x 3          & 256               &                 & (1, 1)          &                   &                  \\
Residual Stack (downsample) + PreLU & 512               &                 & (2, 2)          &                   &                  \\
Residual Stack + PreLU x 3          & 512               &                 & (1, 1)          &                   &                  \\
AdaptiveAvgPool 2D        & -                 & -               & -               & -                 & -                \\ \bottomrule
\end{tabular}
\end{table}

%%%%%%%%%%%%%%%%%%%%%%%%%%%%%%%%%%%%%%%%%%%%%%%%%%%%%%%%%%%%%%%%%%%%%%%%%%%%%%
\clearpage
\subsection{Architectural details - Raw waveform models}

\begin{figure}[h]
  \includegraphics[width=13cm]{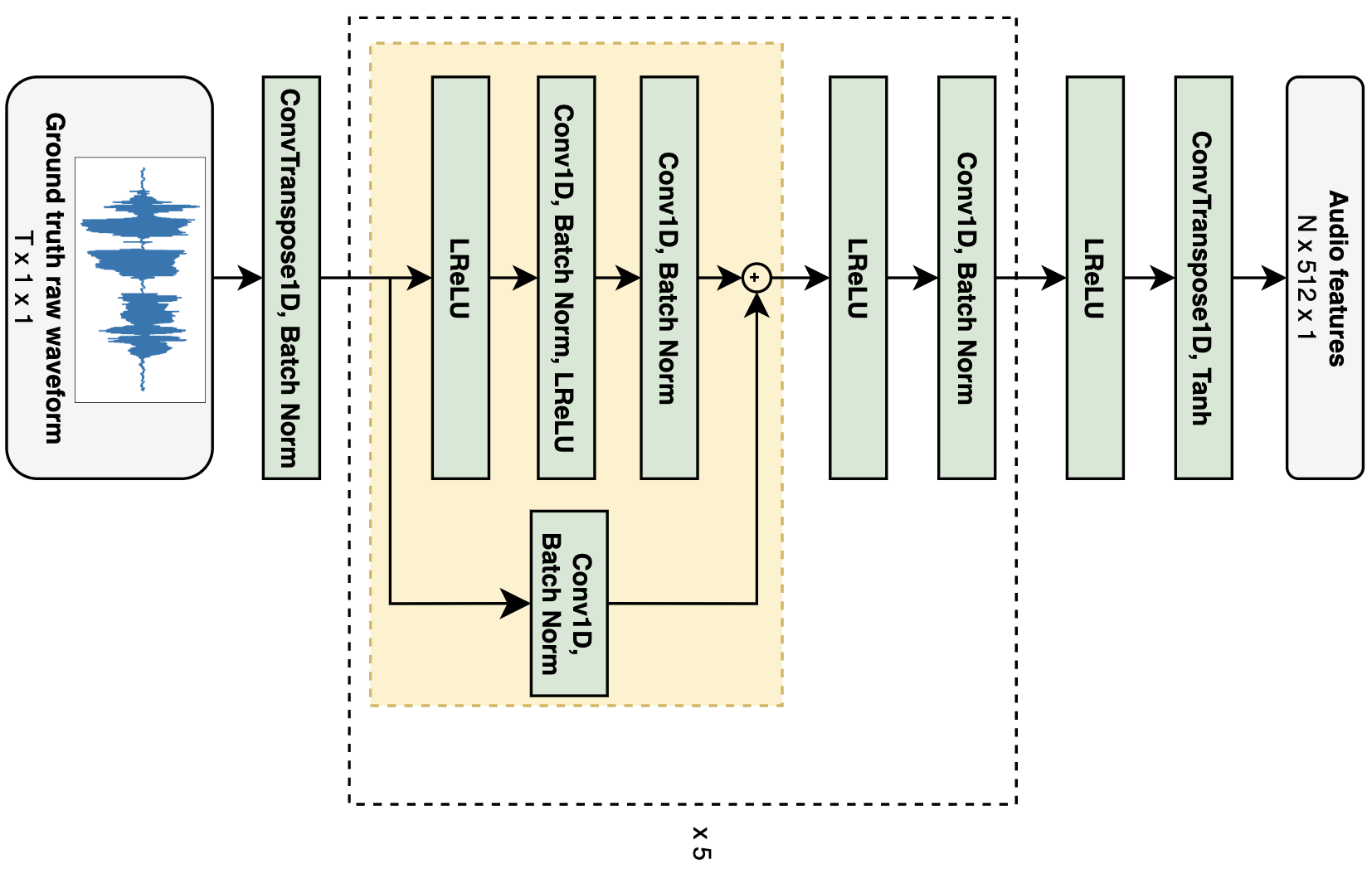}
\centering  
\caption{Raw waveform encoder. Residual stacks are shaded in yellow.}
\label{raw_waveform_encoder}
\end{figure}

\begin{figure}[h]
  \includegraphics[width=18cm]{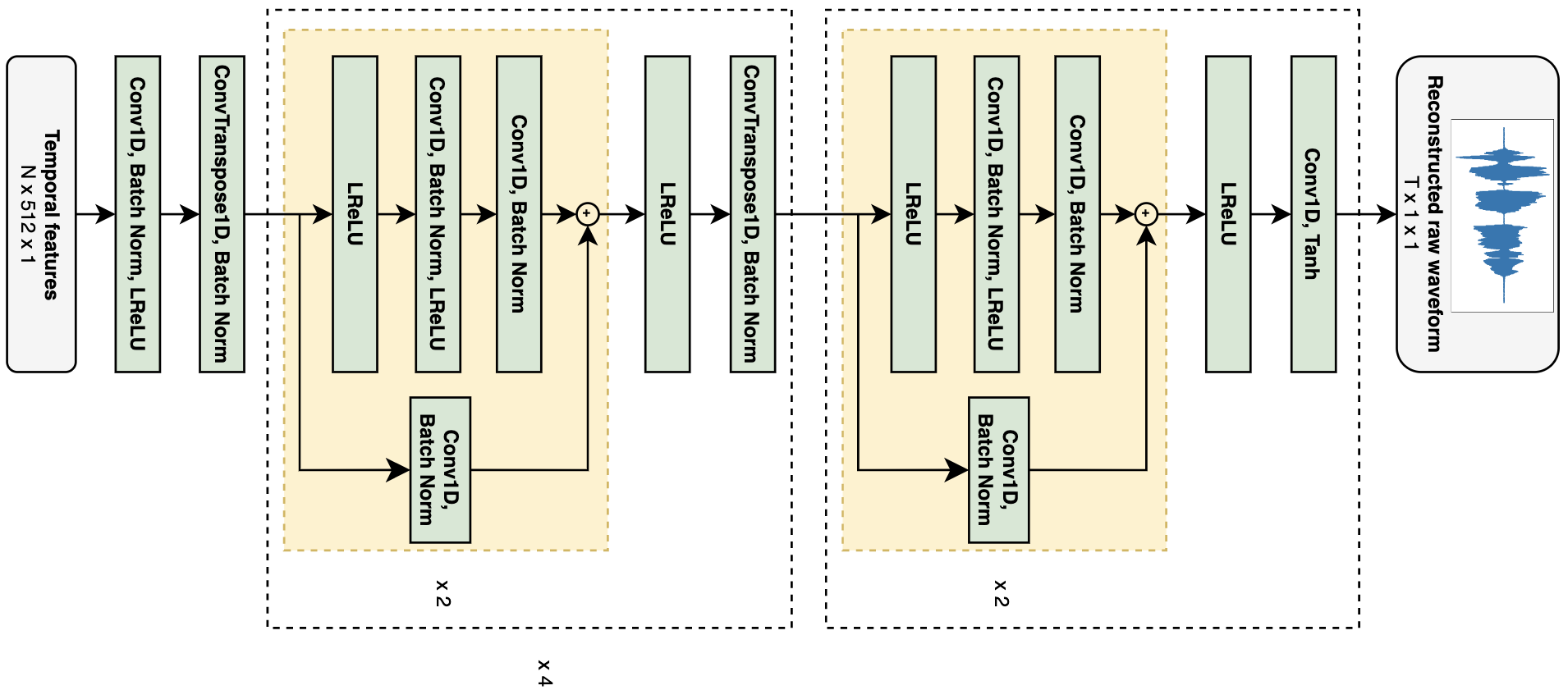}
\caption{Raw waveform decoder. Residual stacks are shaded in yellow.}
\label{raw_waveform_decoder}
\end{figure}

\begin{table*}[h]
\centering
\captionsetup{justification=centering}
\caption{Architecture of residual stack of the raw waveform encoder/decoder with $C$ channels and dilation $D$}
\begin{tabular}{@{}lcccc@{}}
\toprule
\textbf{Layer}              & \textbf{Channels} & \textbf{Kernel} & \textbf{Stride} & \textbf{Dilation} \\ \midrule
LReLU                       &                   &                 &                 &                   \\
Reflection Pad (1, 1)       &                   &                 &                 &                   \\
Conv 1D + BatchNorm 1D + LReLU & $C$                 & 3               & 1               & $D$                 \\
Conv 1D + BatchNorm 1D         & $C$                 & 1               & 1               & 1                 \\
                            &                   &                 &                 &                   \\
\textbf{Skip layer}         &                   &                 &                 &                   \\
Conv 1D + BatchNorm 1D         & $C$                 & 1               & 1               & 1                 \\ \bottomrule
\end{tabular}
\end{table*}

\begin{table*}[h]
\centering
\captionsetup{justification=centering}
\caption{Architecture of raw waveform encoder}
\begin{tabular}{@{}lccccc@{}}
\toprule
\textbf{Layer}              & \textbf{Channels} & \textbf{Kernel} & \textbf{Stride} & \textbf{Dilation} & \textbf{Padding} \\ \midrule
ConvT 1D + BatchNorm 1D        & 32                & 7               & 1               & 1                 & 3                \\
Residual Stack              & 32                &                 &                 & 1                 & -                \\
LReLU + Conv 1D + BatchNorm 1D & 64                & 6               & 3               & 1                 & 2                \\
Residual Stack              & 64                &                 &                 & 1                 & -                \\
LReLU + Conv 1D+ BatchNorm 1D & 128               & 8               & 4               & 1                 & 2                \\
Residual Stack              & 128               &                 &                 & 1                 & -                \\
LReLU + Conv 1D + BatchNorm 1D & 256               & 8               & 4               & 1                 & 2                \\
Residual Stack              & 256               &                 &                 & 1                 & -                \\
LReLU + Conv 1D + BatchNorm 1D & 512               & 8               & 4               & 1                 & 2                \\
Residual Stack              & 512               &                 &                 & 1                 & -                \\
LReLU + Conv 1D + BatchNorm 1D & 1024              & 10              & 5               & 1                 & 3                \\
LReLU                       & -                 & -               & -               & -                 & -                \\
ConvT 1D + Tanh             & 512               & 7               & 1               & 1                 & 3                \\ \bottomrule
\end{tabular}
\end{table*}

\begin{table*}[h]
\centering
\captionsetup{justification=centering}
\caption{Architecture of raw waveform decoder}
\begin{tabular}{@{}lcccccc@{}}
\toprule
\textbf{Layer}                & \textbf{Channels} & \textbf{Kernel} & \textbf{Stride} & \textbf{Dilation} & \textbf{Padding} & \textbf{Output padding} \\ \midrule
Reflection Pad (3, 3)         & -                 & -               & -               & -                 & -                & -                       \\
Conv 1D + BatchNorm 1D + LReLU          & 2048              & 7               & 1               & 1                 & -                & -                       \\
ConvT 1D + BatchNorm 1D                  & 1024              & 10              & 5               & 1                 & 3                & 1                       \\
Residual Stack                & 1024              &                 &                 & 1                 & -                & -                       \\
Residual Stack                & 1024              &                 &                 & 3                 & -                & -                       \\
LReLU + ConvT 1D + BatchNorm 1D         & 512               & 8               & 4               & 1                 & 2                & -                       \\
Residual Stack                & 512               &                 &                 & 1                 & -                & -                       \\
Residual Stack                & 512               &                 &                 & 3                 & -                & -                       \\
LReLU + ConvT 1D + BatchNorm 1D         & 256               & 8               & 4               & 1                 & 2                & -                       \\
Residual Stack                & 256               &                 &                 & 1                 & -                & -                       \\
Residual Stack                & 256               &                 &                 & 3                 & -                & -                       \\
LReLU + ConvT 1D + BatchNorm 1D         & 128               & 8               & 4               & 1                 & 2                & -                       \\
Residual Stack                & 128               &                 &                 & 1                 & -                & -                       \\
Residual Stack                & 128               &                 &                 & 3                 & -                & -                       \\
LReLU + ConvT 1D + BatchNorm 1D         & 64                & 6               & 3               & 1                 & 2                & 1                       \\
Residual Stack                & 64                &                 &                 & 1                 & -                & -                       \\
Residual Stack                & 64                &                 &                 & 3                 & -                & -                       \\
LReLU + Reflection Pad (3, 3) & -                 & -               & -               & -                 & -                & -                       \\
Conv 1D + Tanh                & 1                 & 7               & 1               & 1                 & -                & -                       \\ \bottomrule
\end{tabular}
\end{table*}

\begin{table*}[h]
\centering
\captionsetup{justification=centering}
\caption{Architecture of raw waveform Discriminator for one resolution of raw audio. The Multi-scale Discriminator employs three such identical discriminators.}
\begin{tabular}{@{}lccccc@{}}
\toprule
\textbf{Layer}               & \textbf{Channels} & \textbf{Kernel} & \textbf{Stride} & \textbf{Padding} & \textbf{Groups} \\ \midrule
Reflection Pad 1D (7, 7)     & -                 & -               & -               & -                & -               \\
Conv 1D  + WeightNorm + LReLU & 16                & 15              & 1               & -                & -               \\
Conv 1D  + WeightNorm + LReLU & 64                & 41              & 4               & 20               & 4               \\
Conv 1D  + WeightNorm + LReLU & 256               & 41              & 4               & 20               & 16              \\
Conv 1D  + WeightNorm + LReLU & 512               & 41              & 4               & 20               & 64              \\
Conv 1D  + WeightNorm + LReLU & 512               & 5               & 1               & 2                & -               \\
Conv 1D + WeightNorm          & 512               & 3               & 1               & 1                & -               \\ \bottomrule
\end{tabular}
\end{table*}

%%%%%%%%%%%%%%%%%%%%%%%%%%%%%%%%%%%%%%%%%%%%%%%%%%%%%%%%%%%%%%%%%%%%%%%%%%%%%%
\clearpage
\subsection{Architectural details - Mel spectrogram models}

\begin{table}[h]
\centering
\captionsetup{justification=centering}
\caption{Feedforward module in Conformer}
\begin{tabular}{@{}lcc@{}}
\toprule
\textbf{Layer}  & \textbf{Input dim.} & \textbf{Output dim.} \\ \midrule
LayerNorm       & 256                 & 256                  \\
Linear          & 256                 & 2048                 \\
Swish           & -                   & -                    \\
Dropout (p = 0.1) & -                   & -                    \\
Linear          & 2048                & 256                  \\
Dropout (p = 0.1) & -                   & -                    \\ \bottomrule
\end{tabular}
\end{table}

\begin{table}[h]
\centering
\captionsetup{justification=centering}
\caption{Multi-headed self-attention module in Conformer}
\begin{tabular}{@{}lcc@{}}
\toprule
\textbf{Layer}      & \textbf{Input dim.} & \textbf{Output dim.} \\ \midrule
Positional encoding & -                   & -                    \\
LayerNorm           & 256                 & 256                  \\
Linear (query)      & 256                 & 256                  \\
Linear (value)      & 256                 & 256                  \\
Linear (positional) & 256                 & 256                  \\
Dropout (p = 0.1)   & -                   & -                    \\
Linear (output)     & 256                 & 256                  \\
Dropout (p = 0.1)   & -                   & -                    \\ \bottomrule
\end{tabular}
\end{table}

\begin{table}[h]
\centering
\captionsetup{justification=centering}
\caption{Conformer convolution module}
\begin{tabular}{@{}lccccc@{}}
\toprule
\textbf{Layer}    & \textbf{Channels} & \textbf{Kernel size} & \textbf{Stride} & \textbf{Padding} & \textbf{Groups} \\ \midrule
LayerNorm         & 256               & -                    & -               & -                & -               \\
Transpose         & -                 & -                    & -               & -                & -               \\
Pointwise Conv 1D & 512               & 1                    & 1               & -                & -               \\
GLU               & -                 & -                    & -               & -                & -               \\
Depthwise Conv 1D & 256               & 31                   & 1               & 15               & 256             \\
BatchNorm 1D      & 256               & -                    & -               & -                & -               \\
Swish             & -                  & -                     & -                & -                 & -                \\
Pointwise Conv 1D & 256               & 1                    & 1               & -                & -               \\
Dropout (p = 0.1) & -                 & -                    & -               & -                & -               \\ \bottomrule
\end{tabular}
\end{table}

\begin{table}[h]
\centering
\captionsetup{justification=centering}
\caption{Architecture of mel spectrogram encoder/decoder with $B$ conformer blocks, $D_I$ input dim. and $D_O$ output dim.}
\begin{tabular}{@{}lcc@{}}
\toprule
\textbf{Layer / Module}            & \textbf{Input dim.} & \textbf{Output dim.} \\ \midrule
Linear                             & $D_I$                 & 256                  \\
Dropout (p = 0.1)                  & -                   & -                    \\
                                   &                     &                      \\
\textbf{Conformer blocks}          &                     &                      \\
\textbf{for i = 1, 2, ..., $\mathbf{B}$:}     &                     &                      \\
\hspace{3mm}Feedforward module                 & 256                 & 256                  \\
\hspace{3mm}Multi-headed self-attention module & 256                 & 256                  \\
\hspace{3mm}Conformer convolution module       & 256                 & 256                  \\
\hspace{3mm}Feedforward module                 & 256                 & 256                  \\
\hspace{3mm}LayerNorm                          & 256                 & 256                  \\
                                   &                     &                      \\
Linear                             & 256                 & $D_O$                   \\ \bottomrule
\end{tabular}
\vfill
\hspace{3mm} \\
Note that $B = 2, D_I = 80, D_O = 768$ for the encoder and $B \in \{2, 6, 12\}, D_I = 768, D_O = 80$ for the decoder.
\end{table}

\clearpage
\clearpage

%\vfill
\end{document}